\documentclass[12pt]{elsarticle}      
\usepackage[english]{babel}
\usepackage{amsmath}
\usepackage[toc,page]{appendix}
\usepackage{amssymb}        
\usepackage{amsthm}         
\usepackage{mathbbol}
\usepackage{graphicx}       
\usepackage{subfigure}      
\usepackage{color}
\usepackage[export]{adjustbox}
\usepackage{algorithm}
\usepackage{algpseudocode}

\usepackage{tabularx, booktabs, multirow}
\usepackage{listings}
\usepackage{color}
\usepackage{subcaption}

\DeclareFixedFont{\ttb}{T1}{txtt}{bx}{n}{12} 
\DeclareFixedFont{\ttm}{T1}{txtt}{m}{n}{12}  

\usepackage{color}
\definecolor{deepblue}{rgb}{0,0,0.7}
\definecolor{deepred}{rgb}{0.6,0,0}
\definecolor{deepgreen}{rgb}{0, 0.55, 0}
\definecolor{deeppurple}{RGB}{50, 0, 200}
\definecolor{red}{rgb}{1,0,0}
\usepackage{listings}

\newcommand\pythonstyle{\lstset{
language=Python,
basicstyle=\ttm\tiny,
otherkeywords={self},             
keywordstyle=\ttb\tiny\color{deepblue},
emph={MyClass,__init__},          
emphstyle=\ttb\tiny\color{deepred},    
stringstyle=\color{deepgreen},
frame=tb,                         
showstringspaces=false ,           %
commentstyle=\color{red},
breaklines=true
}}

\lstnewenvironment{python}[1][]
{
\pythonstyle
\lstset{#1}
}
{}

\newcommand\pythoninline[1]{{\pythonstyle\lstinline!#1!}}

\newcommand{\grad}{\nabla}

\newcommand{\bluecom}[1]{{\color{black} {#1}}}

\newcommand{\purplecom}[1]{{\color{deeppurple} {#1}}}

\usepackage{tikz}

\begin{document}

\begin{frontmatter}

\title{A crack-length control technique for phase field fracture in FFT homogenization}
\author[upm,usach]{Pedro Aranda}

\author[upm,imdea]{Javier Segurado}\corref{cor1}
\ead{javier.segurado@upm.es}

\address[upm]{Department of Materials Science, Universidad Politécnica de Madrid, C/ Profesor Aranguren 3, 28040 - Madrid, Spain}
\address[usach]{Department of Mechanical Engineering, University of Santiago de Chile, USACH, C/ Bernardo O’Higgins 3363, Santiago de Chile 9170022, Chile}
\address[imdea]{IMDEA Materials Institute, 28906, Getafe, Madrid, Spain}

\begin{abstract}
Modeling the propagation of cracks at the microscopic level is fundamental to understand the effect of the microstructure on the fracture process. Nevertheless, microscopic propagation is often unstable and when using phase field fracture poor convergence is found or, in the case of using staggered algorithms, leads to the presence of \emph{jumps} in the evolution of the cracks. In this work, a novel method is proposed to perform micromechanical simulations with phase field fracture imposing monotonic increases of crack length and allowing the use of monolithic implementations, being able to resolve all the snap-backs during the unstable propagation phases. The method is derived for FFT based solvers in order to exploit its very high numerical performance in micromechanical problems, but an equivalent method is also developed for Finite Elements (FE) \bluecom{showing the equivalence of both implementations}. It is shown that the stress-strain curves and the crack paths obtained using the crack control method are superposed in stable propagation regimes to those obtained using strain control with a staggered scheme. J-integral calculations confirm that during the propagation process in the crack control method, the energy release rate remains constant and equal to an effective fracture energy that has been determined as function of the discretization for FFT simulations. Finally, to show the potential of the method, the technique is applied to simulate crack propagation through the microstructure of composites and porous materials \bluecom{providing an estimation of the effective fracture toughness}.

\end{abstract}
\end{frontmatter}


\section{Introduction}\label{sec:INTRO}
There is a strong interest on studying fracture of heterogeneous materials at the microscale in order to  understand the effect of the microstructure on the fracture processes at microscale and macroscale. \bluecom{These microscopic simulations are fundamental for the estimation of effective properties in heterogeneous materials, which relate the microstructure with the macroscopic crack propagation \cite{hossain2014effective, schneider2020fft}.}
Many studies can be found focused on microscopic fracture simulations in different heterogeneous materials including concrete \cite{ZHANG2020103567,HUANG2021107762}, biological materials \cite{ABDELWAHAB2012128,GUSTAFSSON2019556}, composites \cite{MISHNAEVSKY20091351,segurado2004new} or polycrystals \cite{ma2021phase,Crocker05}. These works are based on the numerical resolution of a mechanical problem at the microscale using a representative volume elements (RVE) of the microstructure, and the simulations can be focused on the behavior of a single material point \bluecom{to extract some macroscopic fracture property} \bluecom{or to be used as the local response in some concurrent multiscale framework }\cite{OLIVER2017560,ARUNACHALA2023115982}.

Different approaches have been used to incorporate degradation at the microscale into micromechanical simulations. Some models rely on discontinuous techniques to model cracks, such as the introduction of cohesive elements \cite{remmers2003cohesive,segurado2004new}, X-FEM \cite{GUSTAFSSON2019556} or embedded strong discontinuities \cite{OLIVER2017560,SANZ2018598}. These techniques require the use of the Finite Element Method (FEM) to solve the mechanical problem, either using the connection between elements or special purpose elements, and their extension to other numerical framework becomes unnatural. Another option to model degradation in micromechanics are continuum damage (CDM) approaches, e.g. Gurson's \cite{GursonContiDuctRupt1977,TVERGAARD1984157,reusch2008nonlocal} or Lemaitrie's \cite{LEMAITRE198531,MAGRI2021113759}, which enable the representation of material damage using continuum fields. As these models do not consider explicitly the crack discontinuity, but continuous damage fields, they can be solved using alternative numerical methods. However, CDM models have limitations in accurately representing actual cracks and are prone to strong mesh dependency and artificial damage localization in bands of zero thickness \cite{JIRASEKNOTES}. To avoid this pathological discretization dependency, non-local versions of CDM models are generally used to regularize the damaged region by the introduction of some material characteristic length, as the so-called gradient damage models \cite{pham2011gradient, PEERLINGS20017723}. These approaches allow a variational formulation since they can be established by energy minimization considering damage and its gradient \cite{marigo2016overview}, yet do not explicitly consider cracks but damaged regions.

An alternative approach sharing properties with the two previous methods is the phase-field fracture model (PFF), which has become a breakthrough in the simulation of fracture. The method was initiated in the seminal work of Francfort and Marigo \cite{FRANCFORTMARIGOJMPS1998} and later improved by Ambrosio-Tortorelli \cite{Ambrosio1990, ambrosio1992approximation, burke2013adaptive} and Miehe \cite{MieheIJNME2010}. This model relies on a variational approach and resembles gradient damage models in many aspects, as in the mathematical definition of energy functional and the use of a characteristic length \cite{de2016gradient}. Nevertheless, the physics behind PFF models is closer to discontinuous fracture approaches, since the model aims at simulating actual cracks and their evolution. For this purpose, the model represents cracks with a continuous field highly concentrated around the actual crack. Its evolution depends directly on the availability of elastic energy to propagate the crack, which depends on the constitutive model and the heterogeneity of the domain. This allows the calculation of multiple and topologically complex fracture patterns with low discretization dependence \cite{onate2017advances, farrell2017linear}.

As gradient CDM, PFF can be implemented in any numerical solver for boundary value problems. In particular, Fast Fourier Transform (FFT) based solvers represent a great alternative to FEM for the implementation of PFF in micromechanical problems. FFT based solvers have the advantage of their high computational efficiency \bluecom{(the cost of the method with the number of degrees of freedom is proportional to $n$ is $nlog(n)$)} and the natural incorporation of periodic boundary conditions, typical of micromechanical simulations. The method is mature nowadays and extensive reviews of its use in micromechanical problems can be found in the literature \cite{schneider2021review, lucarini2021fft}. Indeed, the PFF model has already been implemented in FFT-based methods. Some examples of are the work of Chen et al. \cite{chen2019fft} or the works of Ernesti et al. \cite{ernesti2019fft,ERNESTI2020112793} which study brittle fracture with fast and memory-efficient FFT-based implicit schemes. 

\bluecom{The application of periodic boundary conditions in problems including fracture or softening is controversial. This is because the crack evolution at the macroscale is a localized phenomena, so the stress-strain behavior of a periodic model considering microgeometries that are infinitely repetitive is not representative of the macroscopic response which includes the effect of damage areas and pristine ones. Nevertheless, periodic boundary conditions can be used when the response of a periodic cell is used as constitutive equation in a multiscale problem \cite{OLIVER2017560,ARUNACHALA2023115982} or when the objective is estimating the effective fracture properties of a heterogeneous material considering clear scale separation, as proposed in the works \cite{hossain2014effective,schneider2020fft}. In these cases, the use of periodic cells with periodic boundary conditions becomes an advantage to minimize cell boundary effects. Focusing on homogenizing the fracture properties, the concept of the effective toughness of an heterogeneous media has been boarded in many works, and presents differences to the homogenization of the elastic properties. This problem can be stated as a path minimization problem in which the calculation of the path with minimum fracture energy is decoupled of the mechanical problem associated with fracture, as shown in the work of Brides et al. \cite{braides1996homogenization}. These ideas are latter developed by \cite{schneider2020fft}, who developed an homogenization solver for the fracture toughness based on a FFT scheme. Alternatively,  Hossein et al. \cite{hossain2014effective} propose a definition of the effective toughness of a heterogeneous microstructures as the maximum of the J-integral during crack propagation at the microscale under some special boundary conditions in finite element simulations (surfing conditions).}

\bluecom{PFF simulations are normally solved incrementally by applying a monotonic increase of some prescribed displacement on the boundary or some component of the macroscopic strain field in the micromechanical case.} This type of control is particularly useful under stable crack growth situations \cite{carpinteri2018multiple}, in which the controlling variable is strictly increasing during the crack propagation. However,  when fracture is unstable ---which implies that the crack propagates rapidly without allowing the material to redistribute stresses or dissipate energy effectively--- controlling with a monotonic displacement/strain present several difficulties. Unstable propagation appears in many macroscopic situations but is the common case for cracks evolving through the microstructure.

The first difficulty when modeling unstable crack growth is that the direct use of monolithic solvers is not possible due to the non-convexity of the minimization problem that depends directly on the Phase-Field formulation \cite{SVOLOS2023104359,gerasimov2016line}.  \bluecom{Several solutions has been proposed to extend the use of modified monolithic algorithms like using line search \cite{gerasimov2016line} or quasi-Newton BFGS schemes \cite{KRISTENSEN2020102446}, which have proven to have good consistency and robustness. Another alternative is the use of staggered numerical schemes \cite{MieheIJNME2010,delorenzis15} --in which the mechanical and damage problems are solved independently, since each of them is convex with respect to its target field \cite{gerasimov2016line}-- which has proven to be a very robust solution. Nevertheless, if any of the above mentioned schemes are used,}
all the crack extension during the unstable crack growth is obtained in a single increment, where crack grows in a \emph{meta-stable} regime, \bluecom{requiring hundreds of iterations in some cases}. Moreover, the actual information about the crack path is missing and system equilibrium can only be found when the crack is highly developed or fully propagated \cite{carpinteri2018multiple}. 

Obtaining the mechanical response during the unstable crack growth requires to follow a snap-back behavior \cite{CARPINTERI1989607}, in which equilibrium points follows a path where the applied displacement or mean strain can decrease.  Since the energy dissipated by the crack must remain strictly increasing during the fracture process to obey the second law of thermodynamics \cite{borden2018phase}, an ideal experiment can be designed in which the applied boundary conditions are readjusted to follow a path in which dissipation is imposed in a monotonic way. Early approaches to this kind of strategies in FEM  consisted of using constraint equations involving local displacements rather than global constrains. As examples of this strategy, in the seminal work of Carpinteri et. al. \cite{CARPINTERI1989607} a Crack Mouth Opening Displacement (CMOD) control scheme is introduced with a cohesive crack model to study \emph{catastrophic softening} in elastic bending simulations. In the works by de Borst \cite{de1987computation} and Rots \cite{rots1987analysis} a modified arc-length method is used to study the snap-back behavior in concrete structures for highly localized failure modes using indirect displacement control or the Crack Mouth Sliding Displacement (CMSD) as constrain respectively. More recently, energy dissipation control was introduced to follow snap-back response in the context of CDM or gradient damage, such as the works of Gutierrez et. al. \cite{gutierrez2004energy} and Verhoosel et. al. \cite{verhoosel2009dissipation} who develop arc-length methods on local external forces or displacements associating them with energy dissipation functionals. An alternative approach was proposed to resolve instabilities in micromechanical problems in composites by Segurado et. al. \cite{segurado2004new}, setting the total CMOD of a set of predefined cracks as the control parameter.  Arc-length based methods have been subsequently extended to capture the snap-back behavior in PFF-FEM  at the macroscopic level \cite{vignollet2014phase,DEBORST201678}. One of the first works to use the PFF crack dissipation energy in an arc-length scheme was Singh et al \cite{singh2016fracture}, who implemented a model with the PFF regularization of the crack as a control parameter. This idea was also exploited later in works such as Bharali et. al. \cite{bharali2022robust}, where monolithic arc-length schemes are compared with local displacement control models in FEM or the work of Zambrano et. al. \cite{zambrano2023arc} where a modified staggered scheme on PFF-FEM with arc-length model is proposed to study the snap-back behavior in brittle fracture. 

All the studies mentioned have been proposed for the FEM and macroscopic simulations (with the exception of \cite{segurado2004new}). To the authors knowledge, no equivalent methodology has been proposed in the context of FFT homogenization solvers, in which crack growth instabilities are very important due to the tortuosity of crack evolution through the microstructure. Moreover, the different arc-length approaches proposed in the context of PFF in FEM \cite{verhoosel2009dissipation,vignollet2014phase,DEBORST201678,singh2016fracture,bharali2022robust,zambrano2023arc} rely on the factorization of extended stiffness matrices and cannot be easily translated to FFT methods which are based on iterative approaches (either fix-point iterations or based on Krylov solvers).

To cover this gap, in this work a general implementation of a phase-field fracture model with energy dissipation control (crack area) is proposed for a micromechanical FFT framework. The strategy proposed relies on a monolithic solver in which strain and fracture phase field fluctuations are the unknowns together with the macroscopic strain. The resulting non-linear problem is solved using Newton-Raphson and includes closed expressions for the linear operators of each iteration to be used in bi-conjugate gradient iterative solvers. The article will first present a brief introduction to PFF models and its implementation in FFT solvers using standard macroscopic strain control. Then, the crack length control technique will be described. The FFT based methodology will be validated against FEM resolution of equivalent problems also controlled with energy dissipation and compared with FFT standard strain control simulations. Finally, \bluecom{calculations on microstructural composites, which include the analysis of the effective toughness as function of the inclusions volume fraction in fiber type composites, is presented together with} complex simulations of 3D fissures and porous materials to show the capabilities of the model.

\section{\bluecom{Theory and models}}\label{sec:MODELS}
Phase field fracture models have been widely developed in the literature \cite{MieheIJNME2010,delorenzis15,onate2017advances}. Those studies deepen in the mathematical and geometrical features of the representation of the crack, mechanical response of the material, the process of fracture and the numerical aspect of the model. The following section provide a brief summary of those aspects to be considered for the PFF model proposed in this work.

\subsection{Phase Field Fracture model}\label{sec:PFF}
\begin{figure}[htbp]
\centering
\includegraphics[width=100mm]{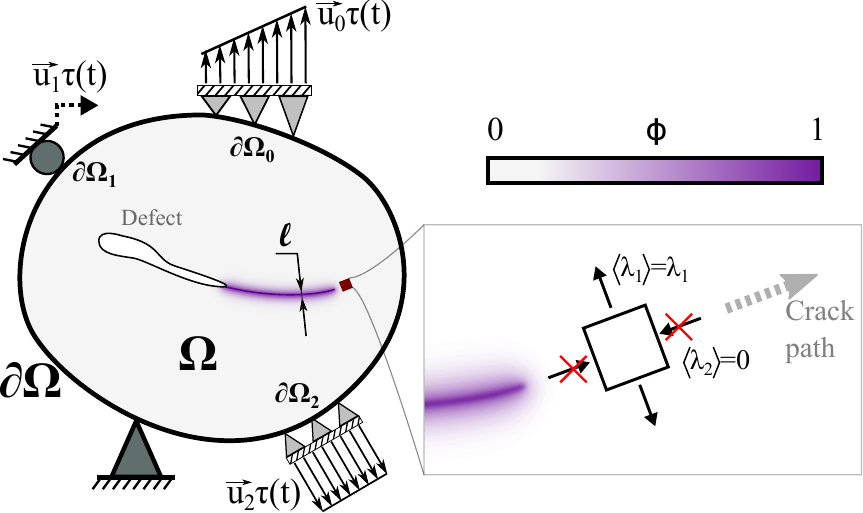}
\caption{\centering{Damage Phase Field representation of the crack.}}
\label{fig1}
\end{figure}

Phase Field fracture model aims at modeling crack growth following an energy criterion. The crack is represented by a continuous scalar field called damage ($\phi$), which varies between 0 (undamaged material) and 1 (fully damaged material). This \emph{Phase Field} allows to represent the crack in a smooth way as it is shown in Fig. \ref{fig1}, evolving spatially and temporally depending on the domain geometry and the material model to which it is coupled.

The problem of the PFF in an elastic media \bluecom{can be stated as a variational problem that involves the displacement field $\vec{u}:\Omega \rightarrow \mathbb{R}^{3}$ and the damage field $\phi:\Omega \rightarrow \mathbb{R}$ from the principle of virtual power \cite{miehe2010phase} 
\begin{equation}
\label{eq:Free_Ener}
\dot{\Psi}(\vec{u},\phi) = \dot{\Pi}_{M} \hspace{3mm}\Rightarrow \hspace{3mm} \dot{E}_{mec}(\vec{u},\phi) + \dot{D}(\phi) - \dot{\Pi}_{M} = 0,
\end{equation}
\noindent where} the functional $\Psi$ represent the internal energy and is composed by the free energy $E_{mec}$ and a fracture dissipation functional $D$, which are balanced by the power of external loads $\Pi_{M}$. The free energy considers the elastic energy of the damaging material
\begin{equation}
E_{mec}= \int_{\Omega} \psi_{e} \mathrm{d}\Omega = \int_{\Omega} g(\phi) \psi_{o}^{+}(\vec{u}) + \psi_{o}^{-}(\vec{u}) \mathrm{d}\Omega,
\label{eq:Emec}
\end{equation}
where the energy density function $\psi_{e}$ is decomposed in its positive and negative parts $\psi_{o}^{+},\psi_{o}^{-}$ \citep{MieheIJNME2010,Ambati2015-1PFF}, that depend on the positive an negative parts of the strain tensor which are obtained from its spectral decomposition
\begin{subequations}
\label{eq:PlusNegPsi}
\begin{eqnarray}
\psi_{o}^{\pm}(\vec{u}) = \frac{1}{2}\lambda \left\langle tr(\boldsymbol{\varepsilon}) \right\rangle^2_{\pm} + \mu \boldsymbol{\varepsilon}^{\pm}(\vec{u}):\boldsymbol{\varepsilon}^{\pm}(\vec{u}) \\
\boldsymbol{\varepsilon }^{\pm} = \left\langle\zeta_{1}\right\rangle_{\pm} \vec{e}_{1}\otimes\vec{e}_{1}+
                              \left\langle\zeta_{2}\right\rangle_{\pm} \vec{e}_{2}\otimes\vec{e}_{2}+
                              \left\langle\zeta_{3}\right\rangle_{\pm} \vec{e}_{3}\otimes\vec{e}_{3},
\end{eqnarray}
\end{subequations}
\noindent where $\boldsymbol{\varepsilon}=sym(\grad{u})$ is the strain tensor, $\lambda$ and $\mu$ are the Lame elastic constant of the material, $\zeta_{i},\vec{e}_{i}$ are the eigenvalues and eigenvectors of the strain and the function $\left\langle\bullet\right\rangle_{\pm}=(\bullet\pm|\bullet|)/2$ represents the positive and negative Heaviside brackets. The positive part of the undamaged elastic energy density functional is degraded with damage to ensure that cracks will not growth in the compressive direction of the principal strain state as is represented in Fig. \ref{fig1}. The degradation is represented with the function $g(\phi)$ of Eq.\eqref{eq:gFunct} that is generally defined with a quadratic behavior,
\begin{equation}
\label{eq:gFunct}
g(\phi) = (1-\phi)^2 - k,
\end{equation}
\noindent where k is a numerical constant that allows to avoid numerical problems when damage is fully developed \cite{MieheIJNME2010}. In this way, the $g(\phi)\psi_{o}^{+}$ functional represent the total elastic energy density available in the system for crack growth. 

In Eq.\eqref{eq:Free_Ener}, \bluecom{the dissipation functional $D$ is only attributed to fracture processes and can be defined by a critical energy release rate $G_c$ related to toughness and the crack density functional} $\gamma_{f}(\phi)$, proposed by Ambrosio and Tortorelli \cite{Ambrosio1990,ambrosio1992approximation}, that represent the geometrical features of the crack and whose integral in the domain defines the total crack surface in 3D or the crack length in 2D
\begin{equation}
\label{eq:CrackLenght}
 \Gamma = \int_{\Omega} \gamma_{f} d\Omega = \int_{\Omega} \frac{1}{2\emph{l}}\phi^{2}+\frac{\emph{l}}{2}\nabla\phi\cdot\nabla\phi d\Omega.
\end{equation}
Together, the functional $D=G_{c}\Gamma$ represents the total energy dissipated due to fracture. 

\subsection{Classical PFF scheme}\label{sec:strain_intro}
\bluecom{With the free energy and dissipation functionals defined, differentiation respect to time and the application of integration by parts allow to derive the strong form of governing equations, defined in the system of Eq.\eqref{eq:OriSystem0} from Eq.\eqref{eq:Free_Ener}}. The domain can be subjected to a set of prescribed tractions and displacement restrictions $\vec{u}_{i}$ on the boundary. For the purpose of this work, we will focus only on applied displacements (Eq.\eqref{eq:OriSystem0c} and Fig. \ref{fig1}), considering a set of surface subdomains $\partial\Omega_{i}$ in which a displacement $\vec{u}_i$ is applied as function of the time.  $\tau(t)$, called sometimes the load factor \cite{bharali2022robust, singh2016fracture}, is used as control parameter to drive the numerical simulation in a proportional way. External work is given by the work of the tractions $\vec{T}$ on the boundary result of the applied displacement, $\Pi_{M}=\int_{\partial\Omega}\vec{T}\dot\vec{u}$,

\begin{subequations}
\label{eq:OriSystem0}
\begin{eqnarray}
\nabla \cdot \left( g \frac{\partial\psi_{o}^{+}}{\partial\boldsymbol{\varepsilon }} + \frac{\partial\psi_{o}^{-}}{\partial\boldsymbol{\varepsilon }}\right) = \nabla \cdot \left( g \boldsymbol{\sigma}_{o}^{+} + \boldsymbol{\sigma}^{-}\right) = \nabla \cdot \boldsymbol{\sigma} = 0 \label{eq:OriSystem0a}\\
g'\psi_{o}^{+}+G_{c} \gamma_{f}'=0 \label{eq:OriSystem0b}\\
\vec{u}=\vec{u_{i}}\tau(t)\hspace{2mm}on\hspace{2mm}\partial\Omega_{i}\hspace{2mm};\hspace{2mm}i=1,2,3... \label{eq:OriSystem0c}
\end{eqnarray}
\end{subequations}

\noindent where $\gamma_{f}'$ is the derivative of the crack functional with respect to $\phi$

\begin{equation}
\label{eq:gamFunct}
\gamma_{f}'(\phi) = \frac{1}{\emph{l}}\phi-\emph{l}\nabla^2\phi.
\end{equation}

The Eq.\eqref{eq:OriSystem0a} represent the mechanical equilibrium equation and Eq.\eqref{eq:OriSystem0b} is a Helmholtz type equation that represent the energy balance that exists between the potential energy and the energy dissipated in fracture \bluecom{that allows to calculate $\phi$ using $\psi_{o}^{+}$ as driving force. Note that, although different phases can be considered, $\phi$ is defined  here in all the domain assuming standard continuity conditions in the displacement and damage fields and jump conditions on their gradients. This assumption allows the damage field to penetrate the interfaces without any restriction depending on the heterogeneity defined for $G_c$ or stiffness. Nevertheless, it is possible to emulate Neumann free boundary conditions of the fracture field in interfaces between two phases using heterogeneous values of \emph{l} between phases, as proposed in \cite{MAGRI2021113759} for FFT based simulations with gradient damage models.}

In micromechanical simulations, for example in FFT homogenization, the domain is a Representative Volume Element (RVE) of the microstructure and its deformation is dictated by the macroscopic strain $\boldsymbol{E}_{M}$. Microscopic strain fields can then be split in two terms
\begin{equation}
\boldsymbol{\varepsilon}(\vec{x}) = \tilde{\boldsymbol{\varepsilon}}(\vec{x}) + \overline{\boldsymbol{\varepsilon}}
\label{eq:split}
\end{equation}
where $\overline{\boldsymbol{\varepsilon}}$ is the average strain in the RVE, that have to be equal to the macroscopic strain $\boldsymbol{E}_{M}$, and $\tilde{\boldsymbol{\varepsilon}}$ is the strain fluctuation field, which average in the RVE vanishes. The external work of Eq.\eqref{eq:Free_Ener} is then expressed as $\Pi_{M}=\boldsymbol{E}_{M}:\boldsymbol{\Sigma}_{M}$ where the macroscopic strain $\boldsymbol{E}_{M}$ and macroscopic stress $\boldsymbol{\Sigma}_{M}$ are defined as volume average of their microscopic counterparts. The macroscopic energy should be equal to the microscopic one, following Hill-Mandel conditions \cite{Lucarini_2021}. Therefore, the restriction Eq.\eqref{eq:OriSystem0c} can be modified. \bluecom{Considering a full strain control, the corresponding PDEs and constrains for the phase-field micromechanical problem are given by Eq.\eqref{eq:OriSystem}.}
\begin{subequations}
\label{eq:OriSystem}
\begin{eqnarray}
\nabla \cdot \boldsymbol{\sigma} = 0 \label{eq:OriSystema}\\
g'\psi_{o}^{+}+G_{c} \gamma_{f}'(\phi)=0 \label{eq:OriSystemb}\\
\boldsymbol{E}_{M} =\frac{1}{\Omega} \int_{\Omega} \boldsymbol{\varepsilon} d\Omega= \tau(t).\label{eq:OriSystemc}
\end{eqnarray}
\end{subequations}

This allows to use the macroscopic strain as a control variable in the solution scheme, by establishing $\tau(t)$ as a linear ramp respect to the time. Note that Eq.\eqref{eq:OriSystemb} could drive to a healing effect of the crack for some damage configurations if the energy $\psi_{o}^{+}$ decrease with time. To overcome this effect, a history function on $\psi_{o}^{+}$ can be implemented as in \cite{miehe2010phase}. The specific numerical implementation for this function and the above described control strategy is detailed in section \ref{ssec:Strain_Contr}.

\subsection{Dissipative control scheme}\label{sec:crack_intro}
Following Griffith theory, if the condition for crack growth of the model corresponds to
\begin{equation}
\left. \frac{-\delta E_{mec}}{ \delta a}\right|_u = G_c
\label{eq:griffith}
\end{equation}
where $E_{mec}$ is the elastic energy of the RVE and $a$ is the crack length, then a crack of a given shape and size will continue growing in a stable manner if
\begin{equation}
\left. \frac{\delta}{\delta a} \left(\frac{-\delta E_{mec}}{ \delta a}\right|_u \right)\leq 0.\label{eq:stable}
\end{equation}
In Phase Field fracture, the terms $E_{mec}$ and $a$ in Eq.\eqref{eq:stable} correspond to the elastic energy functional (Eq.\eqref{eq:Emec}) and the crack area functional (Eq.\eqref{eq:CrackLenght}), respectively. If at a time $t$ Eq.\eqref{eq:stable} is not fulfilled, in order to maintain the fracture condition in Eq.\eqref{eq:griffith}, the  macroscopic strain should be reduced and therefore $\tau(t)$ function of Eq.\eqref{eq:OriSystemc} should decrease with time. As a consequence the mean strain control scheme, in which $\tau(t)$ is imposed as a monotonic function (e.g. a ramp), will result into an uncontrolled crack growth, leading to the problems described in section \ref{sec:INTRO}. 

To overpass this problem, the energy function represented with $G_{c}\Gamma$ in Eq.\eqref{eq:CrackLenght} can be prescribed as function of time, to derive an algorithm  similar to \cite{singh2016fracture} to control the micromechanical simulations by dissipation. In this form, by imposing that the dissipated energy (and consequently the crack length), the fields $\boldsymbol{\varepsilon}$ and $\phi$ fields can be solved. This means that the macroscopic strain $\mathbf{E}_{M}$, used before as simulation control, now it is also an unknown. 

Let the macroscopic strain be written as $\boldsymbol{E}_M=\boldsymbol{f}E$, being $E$ its scalar modulus, that can be understood as a load factor, and $\boldsymbol{f}$ a dimensionless tensor that allows to define the proportionality between the components of the mean strain tensor. Then, the tensor $\boldsymbol{E}_{M}$ is now considered an unknown defined by the scalar load control $E$ and which is equal to the mean field of $\boldsymbol{\varepsilon}$. \bluecom{Alternative mixed boundary conditions could be introduced by using $E$ as the only macroscopic strain controlled component, letting the other components to be determined by macroscopic stress conditions. This can be done in FFT following \cite{Lucarini_2021}.} The new unknown in the system is compensated by adding a new constraint equation, given in Eq.\eqref{eq:SystemDis}

\begin{subequations}
\label{eq:SystemDis}
\begin{eqnarray}
\nabla \cdot \left(\mathbb{C}:\widetilde{\boldsymbol{\varepsilon}} + \mathbb{C}:\boldsymbol{f}E \right)= 0 \label{eq:SystemDisa}\\
g'\psi_{o}^{+}+G_{c} \gamma_{f}'(\phi)=0 \label{eq:SystemDisb}\\
\int_{\Omega} G_{c} \gamma_{f} d\Omega = \tau(t).\label{eq:SystemDisc}
\end{eqnarray}
\end{subequations}

Note that Eq.\eqref{eq:SystemDisc} serves as control equation and is equivalent to the path-following constraint developed in \cite{singh2016fracture} for FEM schemes. The treatment of this equation to solve the problem in a FFT scheme is discussed in section \ref{ssec:Crack_Contr}. Note also that Eq.\eqref{eq:OriSystema} is substituted with Eq.\eqref{eq:SystemDisa} where the stress is replaced by the derivative of the elastic energy density, $\boldsymbol{\sigma}=\mathbb{C}:(\widetilde{\boldsymbol{\varepsilon}}+\boldsymbol{f}E)$. The $\mathbb{C}$ tensor represent the actual elastic stiffness of the material and is calculated with the positive/negative definition of the stress in Eq.\eqref{eq:OriSystem0a} as in \cite{gerasimov2016line}: 

\begin{equation}
\label{eq:C_Calc}
\mathbb{C} = g(\phi) \frac{\partial^{2} \psi_{o}^{+}}{\partial \boldsymbol{\varepsilon}^{2}} + \frac{\partial^{2} \psi_{o}^{-}}{\partial \boldsymbol{\varepsilon}^{2}}.
\end{equation}

The function $\tau(t)$ is established as a linear ramp respect to the time, which implies that in every time a constant crack growth rate will be imposed. Again, Eq.\eqref{eq:SystemDisb} could drive to crack healing. The specific numerical implementation of the control strategy is detailed in section \ref{ssec:Crack_Contr} and include the definition of the history.

\subsection{Effective toughness from micromechanical simulations}\label{ssec:effective_Gc}
\bluecom{
As it was discussed in the introduction, one application of homogenization schemes is the estimation of effective fracture toughness $G_c$ in heterogeneous. The idea behind this concept is to calculate a geometrically and constitutively adequate value, $G_{Ceff}$, for the local toughness of a material that fractures at the macroscopic scale, a value that must be determined from simulations including the microstructure and elastic and fracture properties of the constituents.

To this aim, first its necessary to note that the result PFF simulations in homogeneous media do not exactly recover the input value of the material toughness, but an \emph{effective numerical toughness} which value was determined by Bourdin et al. \cite{bourdin2008variational} for FEM schemes. In this work it was reported that, for a representation in FEM with element size $h$ and a finite value for $\emph{l}$, an overestimation of $G_c$ in Eq.\eqref{eq:CrackLenght} is inherently made despite the $\Gamma$-convergence of the PFF crack in a continuous domain. This may affect the behavior of the mechanical response and change the fracture stress and strain. To be consistent, the closed expression provided by \cite{bourdin2008variational} for an effective numerical toughness $G_{c,num}$ in FEM simulations, detailed in Eq.\eqref{eq:BourdinGc} has to be considered,
\begin{equation}
\label{eq:BourdinGc}
G_{c_num} = G_{c} \left(1+\frac{h}{2\emph{l}} \right).
\end{equation}
This estimation has proven to be accurate in many works for FEM \cite{bourdin2008variational, hossain2014effective, tanne2018crack, costa2023formulations}, but no equivalent expression exists for FFT. An equivalent correction is derived numerically in the Appendix \ref{EffGcsec} for the FFT formulation of this work. 

Once determined the effective numerical toughness of the homogeneous material, the homogenized toughness of an heterogeneous media can be computed using PFF simulations. This homogenized value $G_{Ceff}$ obtained from a microstructural domain would characterize the toughness of a single point in the macroscopic scale, which load state i

As summarized in the introduction, two kind of estimations of the effective toughness can be mentioned. The first one, developed in \cite{hossain2014effective, brach2019anisotropy}, allows to obtain $G_{Ceff}$ from microstructural simulations where the passage of a macroscopic crack is emulated by means of a time depending displacement distribution imposed as boundary condition ('Surfing' boundary condition). The energy release rate is then measured using a J-integral, which contour was defined in a surrounding material with the microstructure effective stiffness (padded material). In this approach, $G_{Ceff}$ is defined as the historical maximum value of the J-integral as the microscopic crack propagates,
\begin{equation}
\label{eq:HosseinGc}
G_{Ceff} = \max_{0 \rightarrow t_f} J(t),
\end{equation}
}
\bluecom{\noindent being $t_f$ the final simulation time.} 

\bluecom{Other alternative has been proposed and used in \cite{schneider2020fft, ernesti2021fast} who develop the postulate of Braides et al. \cite{braides1996homogenization}. The idea is that for certain scenarios the homogenized form of the macroscopic internal energy fulfills
\begin{equation}
\label{eq:SchneiderFunc}
\Psi = \int_{\Omega} \frac{1}{2}\boldsymbol{\varepsilon}:\mathbb{C}_{eff}:\boldsymbol{\varepsilon}d\Omega +
\int_{S_\phi} G_{Ceff}(\vec{t_c}) d S_\phi,
\end{equation}
\noindent with $G_{Ceff}$ the effective toughness,  $\mathbb{C}_{eff}$ the effective stiffness, $S_\phi$ the geometrical surface where the crack is located, and $\vec{t_c}$ a vector defining the normal to the macroscopic crack. With the definition of the homogenized energy in Eq.\eqref{eq:SchneiderFunc}, $G_{Ceff}(\vec{t_c})$ becomes independent of the local strain field. This independence and the definition of an RVE whose dimensions tend to infinity allows to obtain the effective toughness in Eq. \ref{eq:SchneiderFunc} by a topological minimization at the microscale aiming to find the minimal energy crack, 
\begin{equation}
\label{eq:SchneiderGc}
G_{Ceff}(\vec{t_c}) = \lim_{L \rightarrow \infty}  \inf_{S_\phi \subseteq Q_L} \frac{1}{\left|Q_L \right|} \int_{Q_L} G \left|\left| \vec{t_c}+\grad{\Phi} \right|\right| d\Omega ,
\end{equation}
\noindent where $G$ is the local toughness, $\Phi$ is a smooth scalar function characterizing the crack (similar to $\phi$ in PFF model) and $Q_L$ is the RVE volume with size $L$ in which the minimization will be done. In \cite{schneider2020fft} the size of $Q_L$ was studied for a fiber reinforced composite and convergence was found, allowing to obtain results for relatively small RVE sizes.

In this work we postulate an alternative method to obtain an effective toughness $G_{Ceff}$ that is a direct outcome of the PFF-FFT and the technique control developed in this paper. This approach presents some common points with the strategy proposed in \cite{schneider2020fft} but other clear differences. The idea is to simulate the propagation of a crack in the RVE under some given macroscopic loading direction $\mathbf{f}$ until complete cracking of the RVE. Since a rigorous interpretation of the original elastic PFF model \cite{miehe2010phase} cannot consider nucleation, an initial crack should be included in the RVE. When the crack crosses completely the periodic domain $\Omega$, the associated elastic energy $E_{mec}$ in Eq.\eqref{eq:Emec} becomes zero and the energy associated with the surface created by the crack is equal to the total external energy $\Pi_M$ of Eq.\eqref{eq:Free_Ener}, characterized by the area under the curve of the resulting macroscopic mechanical variables $\boldsymbol{\Sigma}_M$ and $\boldsymbol{E}_M$:
\begin{equation}
\label{eq:JaviGc}
G_{Ceff}(\boldsymbol{f}) = \frac{1}{L_{g}}\int_{\Omega} G_c(x)\gamma_f(x) dx = \frac{1}{L_{g}}\int_{E=0}^{E=E_f} \boldsymbol{\Sigma}_M:\boldsymbol{f} dE,
\end{equation}
\noindent where $L_{g}$ is the length of the domain in the direction of the macroscopic crack and $E_{f}$ is the value of the unknown strain $E$ at the crack coalescence. This direction can be defined as equal to the direction of a crack obtained in a homogeneous domain with the same dimensions as $\Omega$ and with the same imposed $\boldsymbol{f}$.

As in \cite{schneider2020fft} the resulting crack is also a result of minimizing a functional depending on a crack field, but in this case the mechanical microfields play a fundamental role. Another difference is that no macroscopic crack direction is explicitly imposed, but result of the loading direction dictated by $\mathbf{f}$. The macroscopic strain direction $\mathbf{f}$ will influence the resulting effective toughness, since both the crack topology and the energy depend on the macroscopic strain tensor $\boldsymbol{E}_M=\boldsymbol{f}E$.}

\section{Numerical implementation}\label{sec:Control}
The following section describes the numerical schemes developed to introduce the aforementioned strain control and crack-length control in FEM and FFT-based schemes.

\subsection{FFT Strain-controlled simulation}\label{ssec:Strain_Contr}
\bluecom{If a macroscopic deformation $\boldsymbol{E}_{M}$ is imposed as the external restriction as in Eq.\eqref{eq:OriSystemc}, the implicit-staggered scheme described in \cite{MieheIJNME2010} can be used to find the fields $\widetilde{\boldsymbol{\varepsilon}}$ and $\phi$ at each time step. Since the functional of Eq.\eqref{eq:Free_Ener} is convex only respect to $\phi$ and $\boldsymbol{\varepsilon}$ separately \cite{gerasimov2016line}, the problem will be split in two, to solve first the Eq.\eqref{eq:OriSystema} considering a constant damage field and then the Eq.\eqref{eq:OriSystemb} considering a constant strain field. In the stagger strategy both sub-problems are solved sequentially until both $\widetilde{\boldsymbol{\varepsilon}}$ and $\phi$ fields converge.

The Fourier-Galerkin method \cite{VONDREJC2014,ZemanGeus2017} is used to solve the micromechanical sub-problem for its robustness and reliability. 
This method is combined with a Krylov approach to solve the Helmholtz equation in the system of Eq.\eqref{eq:OriSystem}. All the numerical framework is implemented in the FFT-based homogenization code FFTMAD \cite{LUCARINI2019, Lucarini_2021}. 

The RVE corresponds to a three dimensional domain $\Omega$ of size $L_1,L_2,L_3$ discretized in a grid of $n_x,n_y,n_z$ voxels, referred to an orthonormal basis ${x,y,z}$. Using the Fourier space, the fields involved in the problem are discretized by trigonometric polynomials defined by a set of frequencies $\vec{\xi}$ and the value of the points of the grid in the real space.

Although the pristine material is linear elastic, the free energy decomposition for a damaged material turns the equilibrium differential equation into non-linear. Consequently, the first sub-problem, called from now on the mechanical problem, will be solved with a Newton-Raphson scheme (NR) using the linearization $\boldsymbol{\sigma} (\Delta \boldsymbol{\varepsilon})\approx\mathbb{C}_{i}: \Delta\boldsymbol{\varepsilon}+\boldsymbol{\sigma}_{i}$, where $i$ is a NR iteration and $\mathbb{C}_{i}$ is the mechanical tangent including damage and evaluated using Eq.\eqref{eq:C_Calc} with the strain field of previous iteration. Focusing on the linearized problem at each Newton iteration,  the strain increment $\Delta \boldsymbol{\varepsilon}$ is the unknown field to obtain an equilibrated linearized stress. The weak form of mechanical equilibrium of Eq.\eqref{eq:OriSystema} is then written as 
\begin{equation}
\label{eq:FouGal_equil0}
 \int_{\Omega} \boldsymbol{\sigma}:\delta \boldsymbol{\varepsilon}  \mathrm{d} \Omega =0
 \hspace{2mm}\Rightarrow\hspace{2mm} \int_{\Omega} \zeta:\left[ \mathbb{G}*\boldsymbol{\sigma} \right] \partial\Omega = 0.
\end{equation}
where the strain test function $\delta\boldsymbol{\varepsilon}$ on the left part of Eq.\eqref{eq:FouGal_equil0} has to be a compatible strain tensor. To enforce this compatibility, a projection operator $\mathbb{G}$ convoluted with a general (non-compatible) second order tensor $\zeta$ is introduced in the weak form \cite{ZemanGeus2017, lucarini2021fft}. Using the properties of the convolution, the resulting form of the weak formulation is expressed in Eq.\eqref{eq:FouGal_equil0}. After proper discretization in trigonometrical polynomials and integration, the test function can be eliminated leading to a discrete linear system of equations in $\mathbb{R}^{6 n_x n_y n_z}$ \cite{VONDREJC2014} which result is the strain increment $\Delta \boldsymbol{\varepsilon}$.}

\begin{equation}
\label{eq:FouGal_equil}
\mathbb{G}*\boldsymbol{\sigma}= \mathbb{G}*\left( \mathbb{C}_{i}:\Delta\boldsymbol{\varepsilon}+\boldsymbol{\sigma}_{i} \right)=0,
\end{equation}

Eq.\eqref{eq:FouGal_equil} corresponds to a linear system of equations on in which the convolution of the stress with the projection $\mathbb{G}$ can be obtained by transformation to Fourier space as
\begin{equation}
\label{eq:LinearOpe_Mech}
\mathcal{F}^{-1}\left\{ \widehat{\mathbb{G}}:\mathcal{F} \left\{\mathbb{C}_{i}:\Delta\widetilde{\boldsymbol{\varepsilon }}_{i+1}\right\} \right\}=
-\mathcal{F}^{-1}\left\{ \widehat{\mathbb{G}}:\mathcal{F} \left\{ \boldsymbol{\sigma}_{i}\right\} \right\},
\end{equation}
\noindent where $\mathcal{F}()$ and $\mathcal{F}^{-1}()$ correspond to the discrete Fourier and inverse Fourier transforms and $ \widehat{ } $ represents a field in Fourier space. The expression of the tensor $\mathbb{G}$ in Fourier space for strain control is given by Eq.\eqref{eq:GFourier} where $\boldsymbol{\xi}$ is the frequency vector. 
\begin{equation}
\label{eq:GFourier}
\widehat{\mathbb{G}}=\mathbb{0}\hspace{2mm}\mathrm{for}\hspace{2mm}\vec{\xi}=\vec{0}\hspace{2mm};\hspace{2mm}\widehat{\mathbb{G}}_{ijkl}=\delta_{ik}\frac{\xi_{j}\xi_{l}}{\vec{\xi}\cdot\vec{\xi}}\hspace{2mm}\mathrm{for}\hspace{1mm}\vec{\xi}\neq \vec{0}.
\end{equation}
\bluecom{In the case of high phase stiffness contrast, as it happens in the presence of a crack, numerical noise might appear. In the present case, to mitigate this noise we use a discrete projection operator based on finite differentiation (in particular the rotated scheme proposed by Willot \cite{willot2015fourier}) which is included using modified frequencies in the Fourier expression of $\mathbb{G}$ (Eq. \ref{eq:GFourier})}

Note that in Eq.\eqref{eq:LinearOpe_Mech} $\Delta\widetilde{\boldsymbol{\varepsilon }}$ is placed instead $\Delta\boldsymbol{\varepsilon }$ since stress $\boldsymbol{\sigma}_{i}$ is obtained by accumulation of $\Delta\widetilde{\boldsymbol{\varepsilon}}$ in the last iteration, as stated in Eq.\eqref{eq:stress_1a}, which considers the prescribed mean strain. \bluecom{The Eq.\eqref{eq:LinearOpe_Mech} is solved using the conjugate gradient method. The left hand side of the equation is a linear operator acting on the discrete field $\Delta\widetilde{\boldsymbol{\varepsilon }}$. The keypoint for the efficiency of the method is that the coefficient matrix does not need to be formed and stored for the iterative process, but only the action of the linear operator on the current value of the strain. }
At the end of each Newton iteration, the stress and the elastic stiffness for the new update $i+1$ are defined in Eq.\eqref{eq:stress_1b} and Eq.\eqref{eq:stress_1c}: 

\begin{subequations}
\label{eq:stress_1}
\begin{eqnarray}
\boldsymbol{\varepsilon}_{i+1}=\boldsymbol{f}E+\sum_{k=0}^{i+1}\Delta\widetilde{\boldsymbol{\varepsilon}}_k \label{eq:stress_1a}\\
\boldsymbol{\sigma}_{i+1}=\boldsymbol{\sigma} \left( \boldsymbol{\varepsilon}_{i+1},^{s}\phi \right) \label{eq:stress_1b}\\
\mathbb{C}_{i+1} = \frac{\partial\boldsymbol{\sigma}}{\partial\boldsymbol{\varepsilon}} \bigg{|}_{\boldsymbol{\varepsilon}_{i+1}}\label{eq:stress_1c}
\end{eqnarray}
\end{subequations}

\noindent where $s$ stands for the last converged stagger iteration. 

Once the mechanical problem is solved, the second sub-problem, called from now on the Helmholtz problem, is solved (Eq.\eqref{eq:OriSystemb}) with a constant strain field, which implies a constant energy field. To impose irreversibility (prevent the healing of an already developed crack), the energy field $\psi_{o}^{+}$ in Eq.\eqref{eq:OriSystemb} 
is usually replaced by some history variable $\mathcal{H}$  \cite{MieheIJNME2010}, leading to
\begin{equation}
\label{eq:Helm2}
g'\mathcal{H}+G_{c} \gamma_{f}'(\phi)=0\rightarrow
\left( \frac{G_{c}}{\emph{l}}+2\mathcal{H} \right)\phi + G_{c}\emph{l} \nabla^2 \phi  = 2\mathcal{H}.
\end{equation}

A common choice for the history is the maximum local positive energy of the complete mechanical process,

\begin{equation}
\label{eq:history}
\mathcal{H} = max(\psi_{o}^{+},\mathcal{H}_{n-1}),
\end{equation}
\noindent where $\mathcal{H}_{n-1}$ stands for the history value obtained in the last converged time and $\psi_{o}^{+}$ is the energy field obtained in the last converged mechanical problem of the actual stagger iteration.

To solve the problem, the linear differential operator on the left hand side of Eq.\eqref{eq:OriSystemb} is replaced by a discrete linear operator in which the Laplacian is computed based on Fourier derivative. \bluecom{Note that in the Fourier transform of Helmholtz equation, standard frequencies are used, contrary to the mechanical case where the rotated scheme is used \cite{willot2015fourier}}.
The resulting linear system of equations to be solved corresponds to
\begin{equation}
\label{eq:LinearOpe_Helm}
\left( \frac{G_{c}}{\emph{l}}+2\mathcal{H} \right)\phi - G_{c}\emph{l}\mathcal{F}^{-1}\left[\vec{\xi}\cdot\vec{\xi}\widehat{\phi}  \right] = 2\mathcal{H}.
\end{equation}

The linear equation in Eq.\eqref{eq:LinearOpe_Helm} is solved by the conjugate gradient method with the discrete $\phi$ field as the unknown. Both equations Eq.\eqref{eq:LinearOpe_Mech} and Eq.\eqref{eq:LinearOpe_Helm} are solved each stagger step $s$ of the implicit staggered scheme of algorithm \ref{alg:StrainCont}, which is based in \cite{MieheIJNME2010}. 

Once the staggered scheme converges and both fields, $\Delta\widetilde{\boldsymbol{\varepsilon }}$ and $\phi$, are obtained for the current time step, the modulus of the macroscopic strain $E$ (control variable) is increased for the next time step. The convergence of each sub-problem and the staggered scheme is determined in relative terms of the unknowns. The mechanical problem error of Eq.\eqref{eq:errors_straina} is defined as the maximum strain increment respect to the field norm and the Helmholtz problem have the characteristic absolute residual of a conjugate gradient method
. The stagger error is based on relative errors of the variables $\widetilde{\boldsymbol{\varepsilon}}$ and $\phi$ as in Eq.\eqref{eq:errors_strainc}:

\begin{subequations}
\label{eq:errors_strain}
\begin{eqnarray}
err_{Mech} = \frac{     max\left(|\Delta\widetilde{\boldsymbol{\varepsilon}}|\right)   }{||\overline{\boldsymbol{\varepsilon}}_{i}||} \label{eq:errors_straina}\\
err_{sta}  = max \left(  \frac{||\boldsymbol{\varepsilon}_{s}-\boldsymbol{\varepsilon}_{s-1}||}{||\boldsymbol{\varepsilon}_{s}||}  , \frac{||\phi_{s}-\phi_{s-1}||}{||\phi_{s}||}\right) , \label{eq:errors_strainc}
\end{eqnarray}
\end{subequations}
where $i$ correspond to a NR iteration, $s$ correspond to a Stagger iteration, $\phi_{j-1}$ correspond with the last iterated field in the CG problem, $|\bullet|$ is the absolute value in every point of a field and $||\bullet||$ is the euclidean norm of a vector or tensor field.

In algorithm \ref{alg:StrainCont} every variable have three index where the upper-right accounts for time, the upper-left accounts for the stagger iterations and the bottom-right accounts for the mechanical NR iterations.
\clearpage

\begin{algorithm}[h]
\small
\begin{algorithmic}[1] 
\State Initial data: $\widetilde{\boldsymbol{\varepsilon}}^{0}=0, E^{0}=0, \phi^{0}=0, \mathcal{H}^{0}=0$
\State Control variable $\Delta E$ fixed
\State Time index: $n=0$ 
\While{$t^{n+1} < t_{final}$} 
\State Initiate Stagger fields: ${^0}\boldsymbol{\varepsilon}=\boldsymbol{\varepsilon}^{n}+\Delta E$\hspace{2mm};\hspace{2mm}${^0}\phi=\phi^{n}$
\State Evaluate constitutive equation: find $^{0}\boldsymbol{\sigma}=\boldsymbol{\sigma}(^{0}\boldsymbol{\varepsilon},^{0}\phi)$ and $^{0}\mathbb{C}=\frac{\partial \boldsymbol{\sigma}}{\partial \boldsymbol{\varepsilon}}\big|_{{^{0}}\boldsymbol{\varepsilon}}$ 
\State Stagger index: $s=0$ 
\While{$Err_{sta} \geq tol_{sta}$} 
\State Initialize mechanical fields: $\boldsymbol{\varepsilon}_{0}={^{s}}\boldsymbol{\varepsilon}$\hspace{2mm};\hspace{2mm}$\boldsymbol{\sigma}_{0}={^{s}}\boldsymbol{\sigma}$\hspace{2mm};\hspace{2mm}$\mathbb{C}_{0}={^{s}}\mathbb{C}$
\State Newton index: $i=0$ 
\While{$Err_{Mech} \geq tol_{Mech}$} 
\State Mech. solver FFT: find $\Delta\widetilde{\boldsymbol{\varepsilon}}\left( \boldsymbol{\sigma}_{i}, \mathbb{C}_{i} \right)$ that fulfills eq. \ref{eq:LinearOpe_Mech}
\State Update strain: $\boldsymbol{\varepsilon}_{i+1}=\boldsymbol{\varepsilon}_{i}+\Delta\widetilde{\boldsymbol{\varepsilon}}$
\State Evaluate constitutive equation: find $\boldsymbol{\sigma},\psi_{o}^{+}$ and $\mathbb{C}$ such:
\State     \hspace{4mm} $\boldsymbol{\sigma}_{i+1}=\boldsymbol{\sigma}(\boldsymbol{\varepsilon}_{i+1},^{s}\phi)$ \hspace{2mm};\hspace{2mm}$\mathbb{C}_{i+1}=\frac{\partial \boldsymbol{\sigma}}{\partial \boldsymbol{\varepsilon}}\big|_{\boldsymbol{\varepsilon}_{i+1}}$ \hspace{2mm};\hspace{2mm} ${\psi_{o}^{+}}_{i+1}=\psi_{o}^{+}(\boldsymbol{\varepsilon}_{i+1})$
\EndWhile ($i=i+1$)
\State     Update fields: ${^{s+1}}\boldsymbol{\varepsilon}=\boldsymbol{\varepsilon}_{i+1}$\hspace{2mm};\hspace{2mm}${^{s+1}}\boldsymbol{\sigma}=\boldsymbol{\sigma}_{i+1}$\hspace{2mm};\hspace{2mm}${^{s+1}}\mathbb{C}=\mathbb{C}_{i+1}$
\State     Update History:  $\mathcal{H}^{n+1}=max\left( {\psi_{o}^{+}}_{i+1},\mathcal{H}^{n} \right)$
\State Helm. solver FFT: find $^{s+1}\phi=\phi\left( \mathcal{H}^{n+1} \right)$ that fulfills eq. \ref{eq:LinearOpe_Helm}
\EndWhile ($s=s+1$)
\State Update strain and damage: $\phi^{n+1}=^{s+1}\phi$ ; $\boldsymbol{\varepsilon}^{n+1}={^{s+1}}\boldsymbol{\varepsilon}$
\EndWhile ($n=n+1$)
\end{algorithmic}
\caption{: Implicit staggered scheme with strain control}
\label{alg:StrainCont}
\end{algorithm}

\subsection{FFT Crack-length controlled simulation}\label{ssec:Crack_Contr}
The crack-length control technique shall allow to find at each time step the fields $\Delta\widetilde{\boldsymbol{\varepsilon }}$ and $\phi$ as well as the macroscopic strain $\mathbf{E}$ for a prescribed value of crack area/length $a=\int_\Omega \gamma_f \mathrm{d}\Omega$. This type of control allows to resolve all the macroscopic stress and strain states during a stable crack propagation, avoiding the large unresolved drops in stress of strain control and resolving the snap-backs, and allows to use a monolithic solver with better convergence \cite{khalil2022generalised}. 

To implement this technique, an FFT-based algorithm has to be derived to solve the non-linear system of Eq.\eqref{eq:SystemDis}. \bluecom{Let $F(q)$ be the non-linear residual for the Newton-Raphson scheme, where $q$ is the set of unknowns $q=\{\widetilde{\boldsymbol{\varepsilon}},\phi,E\}$, and which can be split in three different terms as,}
\begin{gather}
\label{eq:System_CrackLength}
F_{1}(\widetilde{\boldsymbol{\varepsilon}},\phi, E) = {\mathbb{G}*\boldsymbol{\sigma}(\widetilde{\boldsymbol{\varepsilon}},\phi,E) }=\mathcal{F}^{-1} \left\{ \widehat{\mathbb{G}} : \mathcal{F}\left[ \mathbb{C}:\widetilde{\boldsymbol{\varepsilon}}+\mathbb{C}:\boldsymbol{f}E \right] \right\} \nonumber \\
F_{2}(\widetilde{\boldsymbol{\varepsilon}},\phi, E) = \left( \frac{G_{c}}{\emph{l}}+2\mathcal{H} \right)\phi - G_{c}\emph{l}\mathcal{F}^{-1}\left[\vec{\xi}\cdot\vec{\xi}\widehat{\phi}  \right] - 2\mathcal{H} \nonumber \\
F_{3}(\phi) = \int_{\Omega} G_{c} \gamma_{f} d\Omega - \tau(t),
\end{gather}
\bluecom{$F_{1}$ corresponds to the residual of equilibrium equation written following Fourier-Galerkin form (Eqs. \ref{eq:FouGal_equil},\ref{eq:LinearOpe_Mech}) with the stiffness defined in the Eq.\eqref{eq:C_Calc}. $F_{2}$ corresponds to the residual in Helmholtz equation (Eq. \ref{eq:LinearOpe_Helm}) with the Laplacian computed as the Fourier transform of the inner product of the gradients of $\gamma_{f}$. $F_{3}$ is the residual which defines the difference between actual crack area/length and the prescribed one, $\tau(t)$, being the term $\gamma_f$ computed using Fourier transforms as,}
\begin{equation}
\label{eq:newGammaFourier}
\gamma_{f} = \frac{1}{2\emph{l}}\phi^{2}+\frac{\emph{l}}{2} \mathcal{F}^{-1}\left(\vec{\xi}\hat{\phi}\right) \cdot \mathcal{F}^{-1}\left(\vec{\xi}\hat{\phi}\right).
\end{equation}
\bluecom{The Newton-Raphson method linearizes the system respect $q$ considering the corresponding infinitesimal variations of the unknowns} $\Delta q=\{\Delta\widetilde{\boldsymbol{\varepsilon}},\Delta\phi,\Delta E \}$ \cite{farrell2017linear, gerasimov2016line}, 

\begin{equation}
\label{eq:PreCrackLength_linearization}
 \begin{pmatrix} \Delta F_{1}\\\Delta F_{2}\\\Delta F_{3} \end{pmatrix}= \begin{pmatrix} \frac{\partial F_{1}}{\partial\widetilde{\boldsymbol{\varepsilon}}}\Delta\widetilde{\boldsymbol{\varepsilon}}+\frac{\partial F_{1}}{\partial\phi}\Delta\phi+\frac{\partial F_{1}}{\partial E}\Delta E\\ \frac{\partial F_{2}}{\partial\widetilde{\boldsymbol{\varepsilon}}}\Delta\widetilde{\boldsymbol{\varepsilon}}+\frac{\partial F_{2}}{\partial\phi}\Delta\phi+\frac{\partial F_{2}}{\partial E}\Delta E\\ \frac{\partial F_{3}}{\partial\phi}\Delta\phi \end{pmatrix}.
\end{equation} 

The result of this derivative is a linear system in  $\mathbb{R}^{7(n_x n_y n_z) + 1}$ given by the set of equations of Eq.\eqref{eq:CrackLength_linearization}
\begin{gather}
 \Delta F_{1}(q,\Delta q) = \mathcal{F}^{-1} \left\{ \widehat{\mathbb{G}} : \mathcal{F}\left[ \mathbb{C}:\Delta\widetilde{\boldsymbol{\varepsilon}}-2(1-\phi)\frac{\partial \psi_{o}^{+}}{\partial \boldsymbol{\varepsilon}} \Delta\phi + \mathbb{C}:\boldsymbol{f}\Delta E \right]  \right\} \nonumber \\ 
 \Delta F_{2}(q,\Delta q) = -2(1-\phi)\frac{\partial \mathcal{H}}{\partial \boldsymbol{\varepsilon}}: \left( \Delta\widetilde{\boldsymbol{\varepsilon}}+\boldsymbol{f}\Delta E \right)+ \left( \frac{G_{c}}{\emph{l}} + 2\mathcal{H} \right) \Delta\phi - G_{c}\emph{l} \mathcal{F}^{-1}\left[\vec{\xi}\cdot\vec{\xi}\widehat{\Delta\phi}  \right] \nonumber \\
 \Delta F_{3}(q,\Delta q) = \int_{\Omega} \frac{G_{c}}{\emph{l}}\phi\Delta\phi - G_{c}\emph{l} \mathcal{F}^{-1}\left[\vec{\xi}\cdot\vec{\xi}\widehat{\phi}  \right] \Delta\phi d\Omega .
 \label{eq:CrackLength_linearization}
\end{gather}
In $\Delta F_{2}$ of Eq.\eqref{eq:CrackLength_linearization} the derivative of the history with respect damage is included. This definition is fundamental to include the history in the monolithic scheme. Since the history is defined with the function maximum, as is stated in Eq.\eqref{eq:history}, its derivative is by definition a piecewise function

\begin{equation}
\frac{\partial \mathcal{H}}{\partial \boldsymbol{\varepsilon}}=
    \begin{cases}
        0 & \text{if } \psi_{o}^{+} \leq \mathcal{H}\\
        \frac{\partial \psi_{o}^{+}}{\partial \boldsymbol{\varepsilon}} & \text{if } \psi_{o}^{+} > \mathcal{H},
    \end{cases}
\label{eq:Delta_history}
\end{equation}
where $\partial \psi_{o}^{+}/\partial \boldsymbol{\varepsilon}$ can be understood as the undamaged positive part of the stress tensor field mentioned in Eq.\eqref{eq:OriSystem0} and defined in \cite{MieheIJNME2010}. 

It can be observed that the introduction of $E$ as unknown in the system in Eq.\eqref{eq:CrackLength_linearization} together with the restriction in $F_3$ makes the system non-symmetric. Therefore, a Krylov method valid for this type of system, has to be used. In our study, we rely on the Bi-conjugate gradient stabilized method for providing an optimal compromise in robustness and efficiency. If $i$ is the iteration number in the Newton-Raphson scheme, the resulting equation to be solved at each iteration is
\begin{equation}
\label{eq:LinearSyst_CrackCont}
\left[ \Delta F (q_{i-1},\Delta q_{i}) \right] = -\{F(q_{i-1},\tau)\},
\end{equation}
where the unknown is $\Delta q_i$ and the set $q_{i-1}$ of the last iteration is obtained from the successive accumulation of the set $\Delta q_{i}$. In Eq.\eqref{eq:LinearSyst_CrackCont} the value of $\tau$ parameter is set as function of time, so it doesn't depend on $i$. The convergence criterion is similar to Eq.\eqref{eq:errors_strain}c, where the maximum of the errors defined in each equation $F$ for every unknown in $q$ is calculated as in Eq.\eqref{eq:errors_crack}:

\begin{equation}
\label{eq:errors_crack}
err  = max \left(  \frac{E_{i}}{||\boldsymbol{\varepsilon}_{i}||}  , \frac{||\Delta\phi_{i}||}{||\phi_{i}||} , \frac{G_{c}\Gamma_{i}-\tau_{i}}{\tau_{i}}\right).
\end{equation}

In the resulting equations, the definitions of $\mathbb{C}$, $\mathbb{G}$ and $\psi_{o}^{+}$ are the ones given in the strain control method \ref{ssec:Strain_Contr}. Additionally, a successive over-relaxation parameter $\alpha$ can be defined as its shown in algorithm \ref{alg:CrackCont}. This strategy is not strictly necessary for the simulations to succeed, but it has been found that for highly developed cracks the use of $\alpha<1$ sometimes improves convergence. The resulting algorithm is given in algorithm \ref{alg:CrackCont} where the complete sequence of resolution is described. 

\clearpage
\begin{algorithm}[h]
\begin{minipage}{\textwidth}
\centering
\small
\begin{algorithmic}[1]
\State Initial data: $\widetilde{\boldsymbol{\varepsilon}}^{0}=0, E^{0}=0, \phi^{0}=0, \mathcal{H}^{0}=0$
\State Control variable $\tau$ fixed
\State Time index $n=0$
\While{$t^{n+1} < t_{final}$} 
\State Initiate NR scheme fields: $\boldsymbol{\varepsilon}_{0}=\boldsymbol{\varepsilon}^{n}$\hspace{2mm};\hspace{2mm}$\phi_{0}=\phi^{n}$
\State Evaluate constitutive equation: find $\boldsymbol{\sigma}_{0}=\boldsymbol{\sigma}(\boldsymbol{\varepsilon}_{0},\phi_{0})$ and $\mathbb{C}_{0}=\frac{\partial \boldsymbol{\sigma}}{\partial \boldsymbol{\varepsilon}}\big|_{\boldsymbol{\varepsilon}_{0}}$ 
\State Time index $i=0$
\While{$Err \geq tol$} 
\State Solver FFT: find $\Delta q\left( \boldsymbol{\sigma}_{i}, \mathbb{C}_{i},\mathcal{H}^{n} \right)$ that fulfills eq. \ref{eq:LinearSyst_CrackCont} 
\State Update variables\footnote[1]{\scriptsize Parameter $\alpha$ stands for the relaxation parameter in case of successive over-relaxation is used.} : $\boldsymbol{\varepsilon}_{i+1}=\boldsymbol{\varepsilon}_{i}+\alpha\Delta\widetilde{\boldsymbol{\varepsilon}}+\alpha\boldsymbol{f} \Delta E$\hspace{2mm};\hspace{2mm}$\phi_{i+1}=\phi_{i}+\alpha\Delta \phi$
\State Constitutive equation: find $\boldsymbol{\sigma},\psi_{o}^{+}$ and $\mathbb{C}$ such
\State     \hspace{4mm} $\boldsymbol{\sigma}_{i+1}=\boldsymbol{\sigma}(\boldsymbol{\varepsilon}_{i},\phi_{i})$ \hspace{2mm};\hspace{2mm}$\mathbb{C}_{i+1}=\frac{\partial \boldsymbol{\sigma}_{i+1}}{\partial \boldsymbol{\varepsilon}_{i}}$ \hspace{2mm};\hspace{2mm} ${\psi_{o}^{+}}_{i+1}=\psi_{o}^{+}(\boldsymbol{\varepsilon}_{i})$
\State     Update History:  $\mathcal{H}^{n+1}=max\left( {\psi_{o}^{+}}_{i+1},\mathcal{H}^{n} \right)$
\EndWhile ($i=i+1$)
\State Update variables: $\boldsymbol{\varepsilon}^{n+1}=\boldsymbol{\varepsilon}_{i+1}$\hspace{2mm};\hspace{2mm}$\phi^{n+1}=\phi_{i+1}$
\EndWhile ($n=n+1$)
\end{algorithmic}
\end{minipage}
\caption{: Monolithic scheme with crack-length control}
\label{alg:CrackCont}
\end{algorithm}

\subsection{FEM implementation of crack-length control}\label{sec:FEM}
A similar formulation for the crack-length control technique is implemented in a FEM code, in this case FENICS \cite{BarattaEtal2023,AlnaesEtal2014}, in which the same functionals defined in section \ref{ssec:Crack_Contr} are used. Nevertheless an implementation based on Lagrange multipliers is adopted to take into account the restrictions in the boundary conditions of the problem.

\begin{figure}[t]
\centering
\includegraphics[width=120mm]{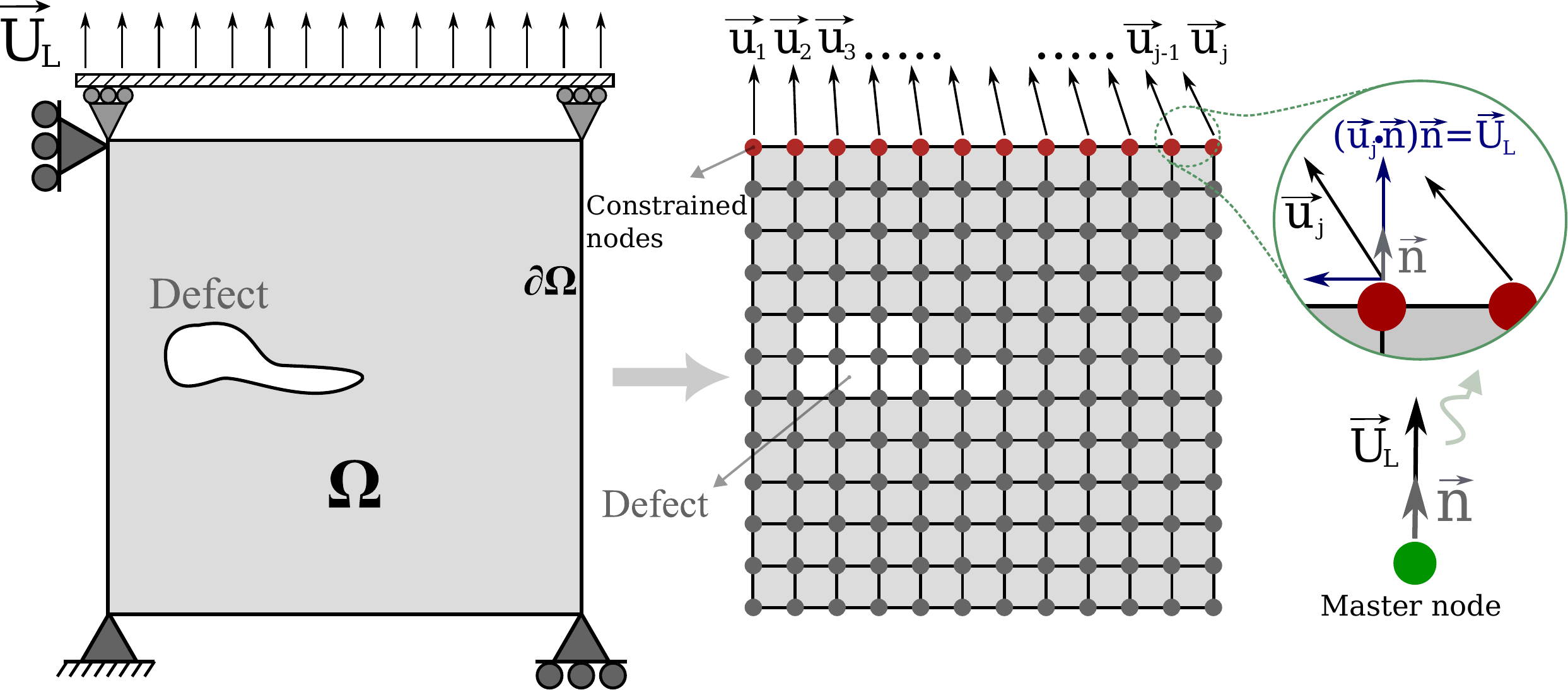}
\caption{\centering{FEM discretization representation and the boundary conditions associated.}}
\label{fig1.5}
\end{figure}

\bluecom{Consider the case on the left of Fig. \ref{fig1.5} of a two-dimensional domain $\Omega$ with boundary $\partial\Omega$ discretized in regular mesh of triangular elements. To set Dirichlet displacements on any boundary in a similar way that in Fig. \ref{fig1} using a proportional factor $\tau(t)$  (Eq.\eqref{eq:OriSystem0c}), the displacement of the nodes involved should be linked by multi-point constraints. Following the scheme developed in \cite{segurado2004new}, a master node $L$ with imposed displacement $\vec{U}_{L}=U_L\vec{n}$ is defined as shown in Fig. \ref{fig1.5}, where $\vec{n}$ is a unit vector in the direction of the applied load. Then, the projection of the displacement in the direction of $\vec{n}$ of the nodes in the moving boundary (constrained nodes in Fig. \ref{fig1.5}) is set equal to the displacement $U_L$ of the master node, making vector $\vec{n}$ to act similarly to the tensor $\boldsymbol{f}$ in the FFT scheme. This multi-point constraint is imposed adding one Lagrange multiplier $\vartheta_{j}$ for each node $j$ in Fig. \ref{fig1.5}.} 

\bluecom{The crack-length restriction of Eq.\eqref{eq:SystemDisc} in FEM is implemented in a slightly different manner respect FFT. The function $\tau(t)$ is also defined as the control parameter but, in order to improve  convergence in the absence of an initial crack field, this value sets a weighted sum of the crack length and the master node displacement $U_{L}$, 
\begin{equation}\label{eq:FEMweights}
    \tau(t) = w_a \int_{\Omega}\gamma_{f}(\phi)\mathrm{d}\Omega+ w_l U_{L}
\end{equation}
being $w_a$ and $w_L$ the constant weights.} Thus, the problem can be solved with crack length control ($w_l=0$), with standard displacement control ($w_a=0$) or with a combination of both. With this restriction and non zero weights, at the beginning of the simulation (when the field $\phi$ is very small) the value of  $\tau(t)$ corresponds almost entirely to the displacement $U_L$. When a crack forms, the crack-length control part will become dominant, allowing decreasing values of $U_L$. It will be shown in the numerical results that this combined control leads to the same results than the simple crack length prescription allowing to represent the snap-back behavior properly. 

\bluecom{To impose the above described control, the FEM scheme developed in \cite{segurado2004new} is adapted to a PFF formulation. The idea behind Eq.\eqref{eq:FEMweights} is to introduce a crack-length control linked to a representative displacement, such as $U_L$, to emulate the FFT scheme in which the formulation is linked to the macroscopic strain. This leads the formulation to be naturally represented by a non-symmetric system. This system can be derived from 
the equations in Eq.\eqref{eq:SystemDisc} including the Lagrange multipliers formulation for the MPC condition described earlier and Eq.\eqref{eq:FEMweights}. The new system is similar to the one obtained in \cite{gerasimov2016line} and is defined by a weak form of a non-linear residual equation $F_{fem}$, which has to be zero for a set of test functions $\Delta q=\{\Delta\vec{u},\Delta\phi,\Delta U_L,\Delta\lambda,\Delta\vartheta_{j}\}$. The weak form of the residual can be split in three terms,
\begin{gather}
F_{1fem}(q,\Delta q) = \int_{\Omega} \boldsymbol{\sigma}(\vec{u}):\boldsymbol{\varepsilon}(\Delta\vec{u}) + g'(\phi)\psi_{o}^{+}(\vec{u})\Delta\phi
+G_{c}\gamma_{f}'(\phi)\Delta\phi\mathrm{d}\Omega \nonumber\\
F_{2fem}(q,\Delta q) = -\sum {}_{j} \left[\vartheta_{j}(\Delta U_L-\Delta\vec{u}_{j}\cdot \vec{n})+\Delta\vartheta_{j}(U_{L}-\vec{u}_{j}\cdot \vec{n})\right]-w_l\lambda\Delta U_{L} \nonumber\\
F_{3fem}(q,\Delta q) =  - \Delta\lambda\left(w_a\int_{\Omega}\gamma_{f}(\phi)\mathrm{d}\Omega+ w_l U_{L}-\tau(t)\right).
\label{eq:LM2}
\end{gather}
$F_{1fem}$ contains the Mechanical/Helmholtz problem, $F_{2fem}$ includes the MPC for displacement on the boundary and a last term which connects $U_L$ with $F_{3fem}$, that correspond to the control functional where Eq.\eqref{eq:FEMweights} is imposed.
The set $\Delta q$ relates to the set of unknowns $q=\{\vec{u},\phi, U_L,\lambda,\vartheta_{j}\}$, where $\vec{u}$, $\phi$ are the displacement and the damage fields respectively and $\vartheta_{j}$ is the set of MPCs Lagrange multipliers. The test function $\Delta\lambda$ is used to introduce the control in the formulation (Eq.\eqref{eq:FEMweights}) and the test function $\Delta U_L$ appears as a consequence of introducing $U_L$ as an extra variable, similar as in the FFT scheme where the control equation is a consequence of introducing the macroscopic strain as an unknown. In this work, the mixed function space of $\vec{u}$ and $\phi$ in $\Omega$ is discretized using triangular elements with bi-linear interpolation and full integration. For $\vartheta_{j}$ and $\lambda$, single valued elements are introduced.

A monolithic Newton scheme is used to solve the weak form of the residual $\mathcal{F}_{fem}=F_{1fem}+F_{2fem}+F_{3fem}$, finding the set of functions $q$ that makes $\mathcal{F}_{fem}=0$ for any suitable function $\Delta q$ and for a prescribed value of $\tau$. To solve this non-linear problem in the Newton-Raphson scheme, the $\mathcal{F}_{fem}$ needs to be linearized to a bilinear form. For this purpose, another set of trial functions $\Delta q_t$ are defined similarly to the test functions $\Delta q$ and used to obtain the Gateaux derivative of the functional of Eq.\eqref{eq:LM2},  $\Delta\mathcal{F}_{fem}$. 
For a Newton iteration $i$, being $q_i$ the current value of the fields, the linear problem to be solved consists in finding the fields $\Delta q_t$ which fulfills
\begin{equation}
\label{eq:LM3}
\left[ \Delta\mathcal{F}_{fem} (q_i,\Delta q,\Delta q_{t}) \right] = -\{\mathcal{F}_{fem}(q_i,\Delta q)\},
\end{equation}
for every possible test function $\Delta q$, leading to the next iteration defining $q_{i+1}=q_i+\Delta q_{t}$}. As in the FFT implementation, the resulting linear system of Eq.\eqref{eq:LM3} is non-symmetric and is solved using a direct LU solver from UMFPACK library. Note that in the variational problem, additional Dirichlet boundary conditions can be imposed. The particular conditions depend on the problem and are specified in section \ref{sec:results}.

\section{Validation and comparison with FEM}\label{sec:results}
The strain and crack-length control schemes are used here to perform a set of simulations that are described in the following section. A first part will include a comparison of FFT and FEM simulations, in order to validate the equivalence of the control method proposed for both frameworks and to analyze the differences between the two type of solvers. Next section will analyze the consequences of the crack-length control developed in terms of Griffith postulates, as well the equivalence of the predicted crack paths obtained using crack control or a standard staggered solved.

Both macroscopic tensile and shear cases are used, and the corresponding control variables are specified in table \ref{table:1}. Three linear elastic materials are used in the simulations, named as matrix, flexible material and rigid inclusion, and their properties are given in table \ref{table:2}. In addition, a material representing the empty space is used for the cases including voids. 

\begin{table}[htbp]
\centering
\begin{tabular}{|c|c|c|} 
\hline
$    $ & Tensile & Shear \\
\hline
Crack-length\hspace{1mm}control & 
\begin{minipage}[b]{2.5cm}
\begin{equation*}
\scriptsize \boldsymbol{f}=\begin{pmatrix} 0&0&0\\0&1&0\\0&0&0 \end{pmatrix}
\end{equation*}
\end{minipage} & 

\begin{minipage}[b]{2.5cm}
\begin{equation*}
\scriptsize \boldsymbol{f}=\begin{pmatrix} 0&0&0\\0&0&1\\0&1&0 \end{pmatrix} 
\end{equation*}
\end{minipage} \\ 
\hline
Strain\hspace{1mm}control & 
\begin{minipage}[b]{3.5cm}
\begin{equation*}
\scriptsize \boldsymbol{E}_{M}=\begin{pmatrix}0&0&0\\0&\tau(t)&0\\0&0&0 \end{pmatrix}
\end{equation*}
\end{minipage} & 
\begin{minipage}[b]{3.5cm}
\begin{equation*}
\scriptsize \boldsymbol{E}_{M}=\begin{pmatrix}0&0&0\\0&0&\tau(t)\\0&\tau(t)&0\\ \end{pmatrix}
\end{equation*}
\end{minipage}\\
\hline
\end{tabular}
\caption{Control tensor features.}
\label{table:1}
\end{table}

\begin{table}[h!]
\centering
\begin{tabular}{|c|c|c|c|c|} 
\hline
$    $ & Matrix & Void & Flexible & Rigid \\
$    $ & $    $ & $  $ & material & Inclusion \\
\hline
$E_{Young} [GPa]$ & 20.8 & 1e-8 & 10.4 & 104 \\
\hline
$\nu$ & 0.3 & 0.4 & 0.3 & 0.3 \\
\hline
$G_{c} [J/m^{2}]$ & 2700 & - & - & - \\
\hline
\end{tabular}
\caption{Mechanical and PFF properties of materials.}
\label{table:2}
\end{table}

Regarding the PFF properties, all cases were simulated with a Gc parameter of $2700 J/m^{2}$. The characteristic length $\emph{l}$ was selected depending on the discretization of each case in order to have a sufficient representation of the function $\phi$ around the cracks. \bluecom{ For each discretization $\emph{l}$ is set as two times the distance between FFT voxels, and this value is used both in FEM and FFT simulations.} This election is made considering a geometrical similarity between a voxel and an element, since this value is reported in \cite{MieheIJNME2010} for PFF-FEM problems as the minimum required for $\Gamma$ functional match the crack-length. In the case of FEM simulations, the weights in Eq.\eqref{eq:FEMweights} are set as $w_l=w_a=1$. In FFT simulations of this section, no relaxation parameter is used ($\alpha=1$ in Algorithm 2).

\subsection{FFT-FEM comparison}\label{sec:TensileFFTval}
\begin{figure}[htbp]
\centering
\includegraphics[width=140mm]{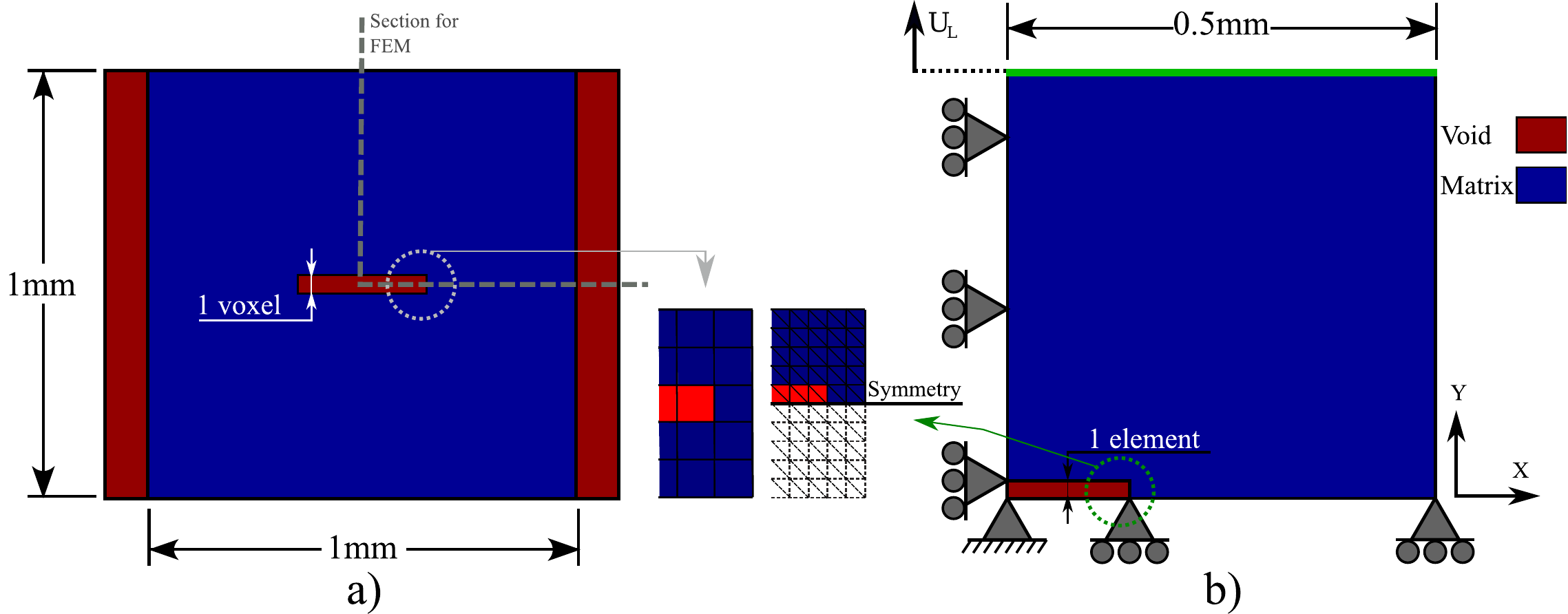}
\caption{\centering{FEM and FFT geometry models.}}
\label{fig2}
\end{figure}

FEM and FFT simulations of the same problem using crack length control are performed using the boundary conditions represented in Fig. \ref{fig2}. In the case of FFT, three-dimensional models with a single voxel in the Z-direction are used, while two-dimensional plane strain condition is used in FEM models.

FEM model is show in Fig. \ref{fig2}b, where two symmetry boundary conditions are applied in axis Y and X on bottom and left edges respectively. A free strain boundary condition is used for the right edge and a displacement restriction for all points in Z direction are applied to represent plain strain. The boundary with nodes which displacement is assigned to be the one of the master node $U_{L}$ described in section \ref{sec:FEM} is remarked in green in Fig. \ref{fig2}b. The initial crack is emulated with the region indicated in the same figure with the properties of the void material of table \ref{table:2}.

\begin{figure}[h]
\centering
\includegraphics[width=120mm]{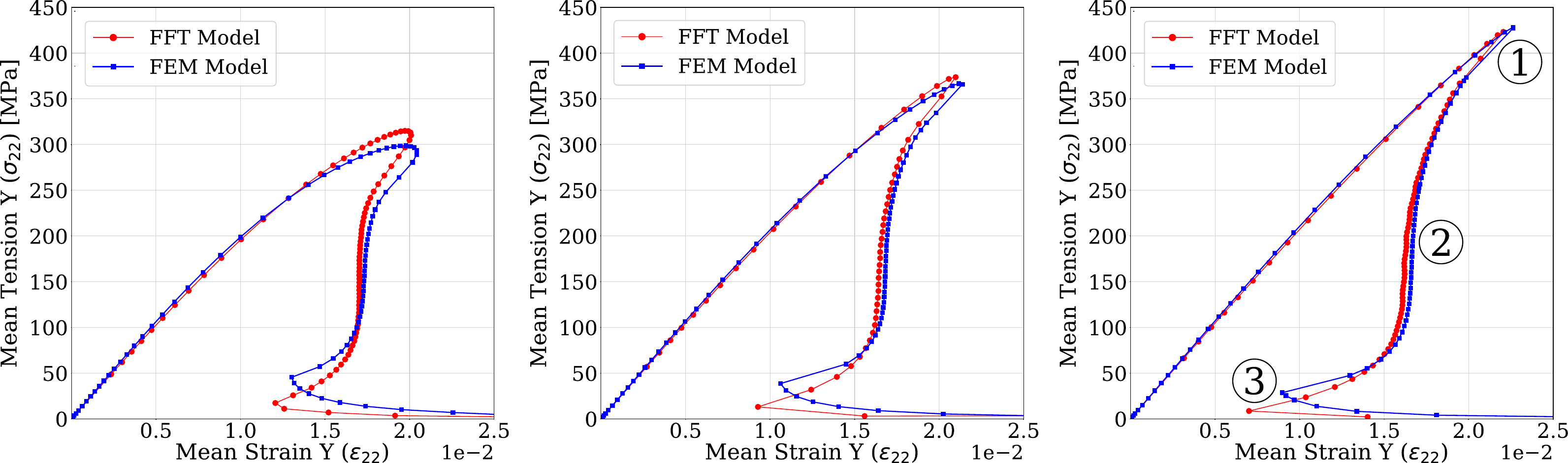}
\caption{\centering{\purplecom{FEM/FFT comparison curves for 3 different discretizations, on the left 39x39 voxels, center 77x77 and on the right 155x155}}}
\label{fig3_a}
\end{figure}

\begin{figure}[h]
\centering
\includegraphics[width=120mm]{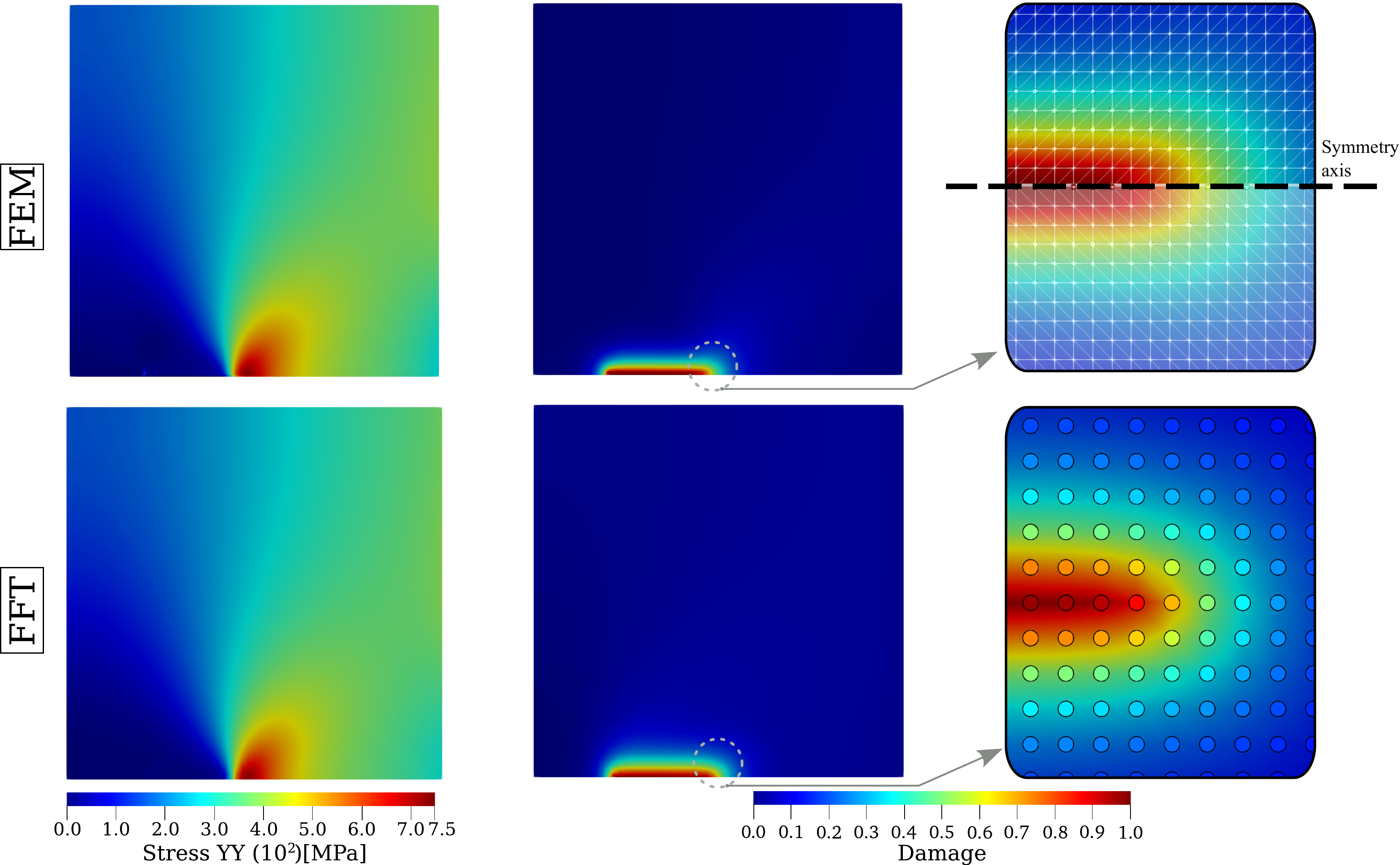}
\caption{\centering{\purplecom{Damage/stress fields for 155x155 voxels case in the second step of Fig. \ref{fig3_a}.}}}
\label{fig3_b}
\end{figure}

In the FFT model of Fig. \ref{fig2}a, no symmetry is applied and periodic boundary conditions are used in the full simulation domain. To represent the free boundary condition in FFT, two bands of void material are placed in left and right edges of the geometry. Considering the load direction $\boldsymbol{f}$ selected, this representation of the free boundary resembles plane strain conditions which are used in the 2D FEM simulations. The thickness of the initial crack in FFT model is one voxel, so to fit the symmetry condition in FEM with the same $\emph{l}$, the number of elements is the same number of voxels but in the half of space (See comparison in Fig. \ref{fig2}). Cases with 39x39 voxels, 77x77 and 155x155 voxels and their equivalent cases in FEM are solved to demonstrate the equivalence of the  control technique in both numerical frameworks. The mechanical response for all cases is described in Fig. \ref{fig3_a} and the damage an stress fields are depicted in Fig. \ref{fig3_b} for the case of 155x155 voxels simulation and his equivalent in FEM for the step $\tikz[baseline=(char.base)]{\node[shape=circle,draw,inner sep=2pt] (char) {2};}$ of Fig. \ref{fig3_a}.

As described in \cite{carpinteri2018multiple}, the fracture process follows three steps, that are stated with numbers in Fig. \ref{fig3_a} for the case of 155x155 voxels. In step $\tikz[baseline=(char.base)]{\node[shape=circle,draw,inner sep=2pt] (char) {1};}$ crack growth starts and the snap-back response is observed after a high stress concentration as is observed in Fig. \ref{fig3_b}. During step $\tikz[baseline=(char.base)]{\node[shape=circle,draw,inner sep=2pt] (char) {2};}$, this stress concentration is kept at the crack tip as is also observed in Fig. \ref{fig3_b}. Step $\tikz[baseline=(char.base)]{\node[shape=circle,draw,inner sep=2pt] (char) {3};}$ represents the end of the fracture process. 

\bluecom{The difference between FEM and FFT results when the number of elements in the FEM model is chosen to be equivalent to the number of FFT voxels can be due to the dissimilar approximations of both techniques, for example in the representation of the fields and integration. Nevertheless, it can be observed in Fig. \ref{fig3_a} that convergence with discretization is reached for both FEM and FFT for relatively coarse meshes. In the case with the finest discretization, 155x155 voxels, step $\tikz[baseline=(char.base)]{\node[shape=circle,draw,inner sep=2pt] (char) {1};}$ and $\tikz[baseline=(char.base)]{\node[shape=circle,draw,inner sep=2pt] (char) {2};}$ are essentially superposed, and only very small differences can be observed in the response for the step $\tikz[baseline=(char.base)]{\node[shape=circle,draw,inner sep=2pt] (char) {3};}$. This small deviation in the final stage, when crack is almost fully developed, can be due to the differences in the particularities of the control technique in both numerical approaches.} 

\subsection{Fracture stability study: Strain and crack control comparison}\label{sec:stability}
As shown in the previous section, a complete and stable representation of the cases that would have an unstable fracture in deformation control has been obtained with crack length control. 

\begin{figure}[t]
\centering
\includegraphics[width=110mm]{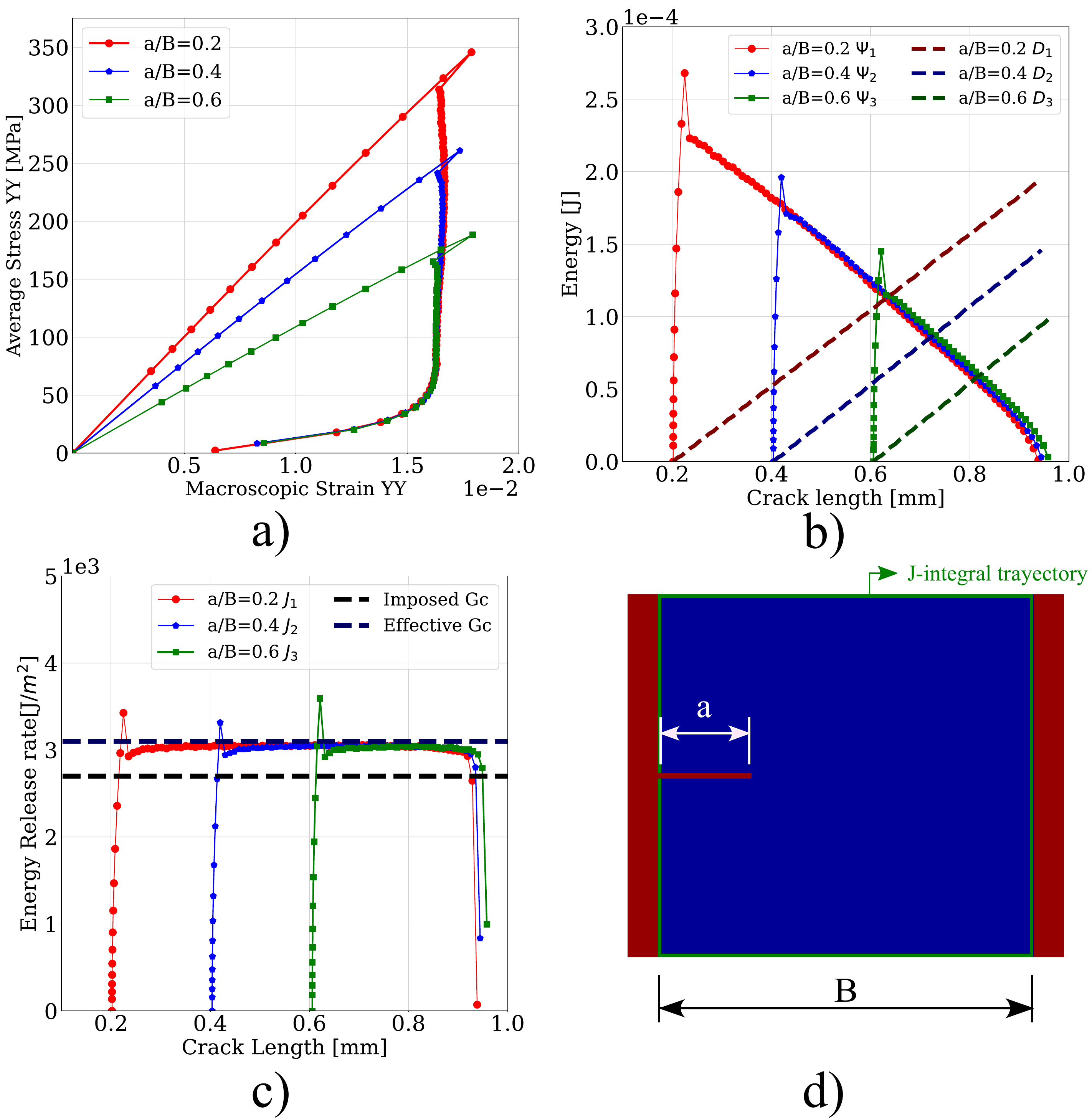}
\caption{\centering{FFT study on the stability of the crack growth in a Griffith panel. a) stress-strain curves, b) Elastic energy degradation as function of crack length, c) J-integral measurements, d) Geometry of the Griffith panel}}
\label{fig4}
\end{figure}

To further analyze the conditions during propagation, simulations of a Griffith panel with initial crack length $a$ and different width $B$ are performed (Fig. \ref{fig4}d). In Fig. \ref{fig4}a the stress-strain response for three $a/B$ ratios on simulations of 257x257 voxels is shown. It can be seen that all curves follows the same stress-strain path once the crack start, so this path correspond to the limit in which the crack propagates, independently of the original crack length. In the classic Griffith fracture theory, these states correspond to the limit in which the elastic energy $G$ release rate reaches the critical energy release energy rate of the material $G_{c}$ (Eq.\eqref{eq:griffith}).

To validate the fulfillment of Griffith condition during propagation, two graphs are represented in Fig. \ref{fig4}b and Fig. \ref{fig4}c. The first one shows the available potential energy of the material $\Psi^{+}$ in continuous lines with points, and the energy dissipation due to crack $D$ using discontinuous lines, both with respect to the crack length. It can be seen that, for same crack length, cases with an initial crack have the same available potential energy that cases in which the crack is already growing. Also, it can be observed the similarity of slopes between dissipation and energy curves, which implies the rates are the same, as dictated by the minimization of the total energy. Nevertheless, to obtain a more accurate measure of $G$, the J-integral is computed, following a method similar to \cite{hossain2014effective}. The trajectory followed to compute the line integral corresponds with the outer contour of the matrix material of Fig. \ref{fig4}d. and the measure of the energy release rate is shown in Fig. \ref{fig4}c for all cases.

\bluecom{As explained in section \ref{ssec:effective_Gc}, it can be observed that the value of the fracture energy obtained is slightly higher than the critical energy release rate used as input data, $G_{c}$. This numerical effective toughness, $G_{Ceff_o}$, depends on the actual material toughness $G_{c}$ and the discretization and characteristic length $l$, and converges to $G_{c}$ with the discretization. This value has been quantified for FFT in the Appendix \ref{EffGcsec}.} It is observed that the value of $G$ obtained with J-integral is very accurate with $G_{Ceff_o}$ computed (error below XX\%) , which is consistent with the idea that this simulation conditions represents a stable fracture process. This value of $G$ is only slightly different to $G_{Ceff_o}$ at the beginning of the process and at the end. The loss of energy release rate shown at the end of the process is an artifact in the measure of J, because at that stage the crack reaches the integral path. The excess of energy and $G$ that is observed in Fig. \ref{fig4}b and Fig. \ref{fig4}c at the beginning of the crack propagation is similar to what has been reported in \cite{hossain2014effective,singh2016fracture}. The reason for this peak may be related to the energy needed to nucleate a phase field crack from a notch or mesh discontinuity. This feature can released using a crack tip enrichment \cite{singh2016fracture,zambrano2023arc} in the definition of the initial crack geometry, but since this effect is only relevant in the first point of propagation standard representation of initial cracks are kept for clearness in this study. \bluecom{Despite the foregoing, this peak in the $J$ integral has a negligible contribution in the dissipated energy (area under the curve in Fig. \ref{fig4}a).

In addition to the $J$ integral, the value of the numerical effective toughness is obtained using the procedure proposed in section \ref{ssec:effective_Gc}, measuring the area under the curve in the simulations of Fig. \ref{fig4}b and dividing it by the crack length developed ($B-a$). Very similar results of $G_c$ are obtained for the three curves, with an average value of $3232.6[J/m^2]$, which correspond to an error below 5\% of the estimated value of $G_{Ceff_o}$, similar accuracy than the one obtained using the J-integral.}

\subsubsection*{Comparison of strain control and crack length control}
\begin{figure}[t]
\centering
\includegraphics[width=100mm]{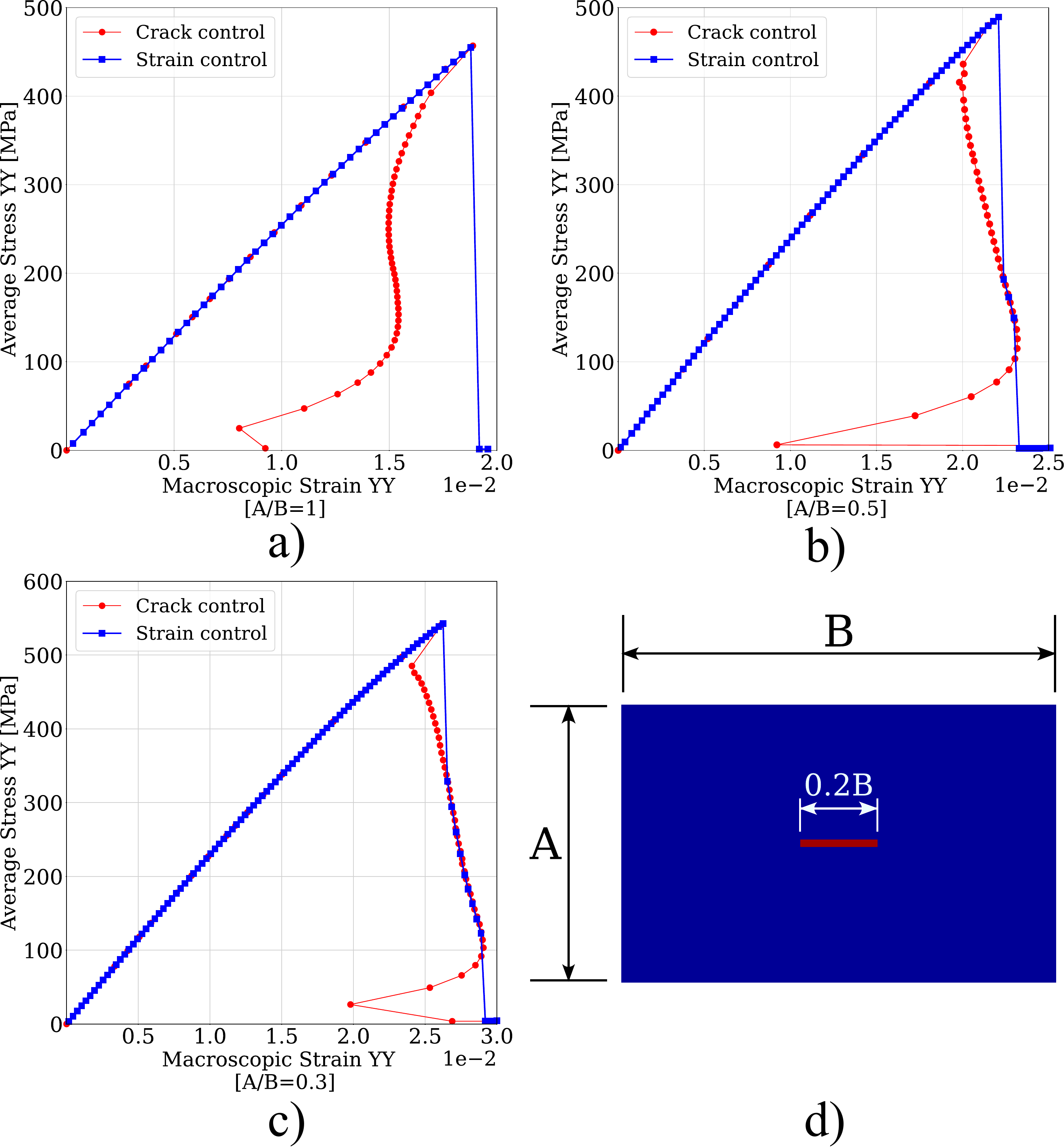}
\caption{\centering{Comparison of Strain/Crack length control for different geometries of Griffith panel, using curves of average strain/stress in direction Y for: a) $A/B=1$ case, b)$A/B=0.5$ case and $A/B=0.3$ case, with dimensions reference in d).}}
\label{fig5}
\end{figure}

Since the stress-strain path for the crack-length controlled simulations has proven to be stable, similar path is expected to be obtained in strain controlled cases where propagation is always stable. This comparison further validates the crack control technique by observing how a strain stable fracture response is represented identically by both models.
The case studied correspond to the same plate in Fig. \ref{fig5}d, with the conditions described in the beginning of section \ref{sec:results} and solved with crack-length control and strain control models. Since this case has demonstrated to be unstable for strain control in section \ref{sec:TensileFFTval}, the ratio $A/B$ of Fig. \ref{fig5}d is modified to fit cases of $A/B=1$, $A/B=0.5$ and $A/B=0.3$ with 1x243x243, 1x121x243 and 1x81x243 voxels correspondingly. The transition from unstable crack propagation to stable is determined by the derivative of the Energy release rate with respect to crack length (Eq. \ref{eq:stable}), and it can be shown using simple linear elastic fracture mechanics that the reduction of the Griffith panel height $A$ reduces the release energy for a differential crack extension leading eventually to a stable regime. Also, since flat displacement on the upper surfaces is equivalent to periodicity for this geometry, this simulation is equivalent to an infinite array of cracks aligned vertically. In this case, reducing the $A/B$ ratio increases the density of periodic fissures respect to the total volume, allowing to redistribute the available elastic energy of the volume in more fracture processes. This causes that rate of dissipation in each fissure decrease, making them more stable. This has been demonstrated in \cite{rooke1976compendium}, where it is explained that the influence of an arrangement of parallel cracks in the perpendicular direction to the crack leads to a decrease of the stress intensity factor in each crack, which implies a generalized decrease of the evolution of the energy release rate. The results obtained are shown in Fig. \ref{fig5}.

It is observed that both types of control leads to the same stress-strain curves during the stable propagation stages. Moreover, the curves obtained using crack-control can be used to determine the range of crack lengths in which growth is stable for a given geometry. Another interesting result is that, in this case, if the staggered approach is used with a  very restrictive tolerance it results in a crack shape equal to the one obtained under crack length control, validating the \emph{metastable} propagation of the crack during the staggered iterations.

\begin{figure}[htbp]
\centering
\includegraphics[width=\textwidth]{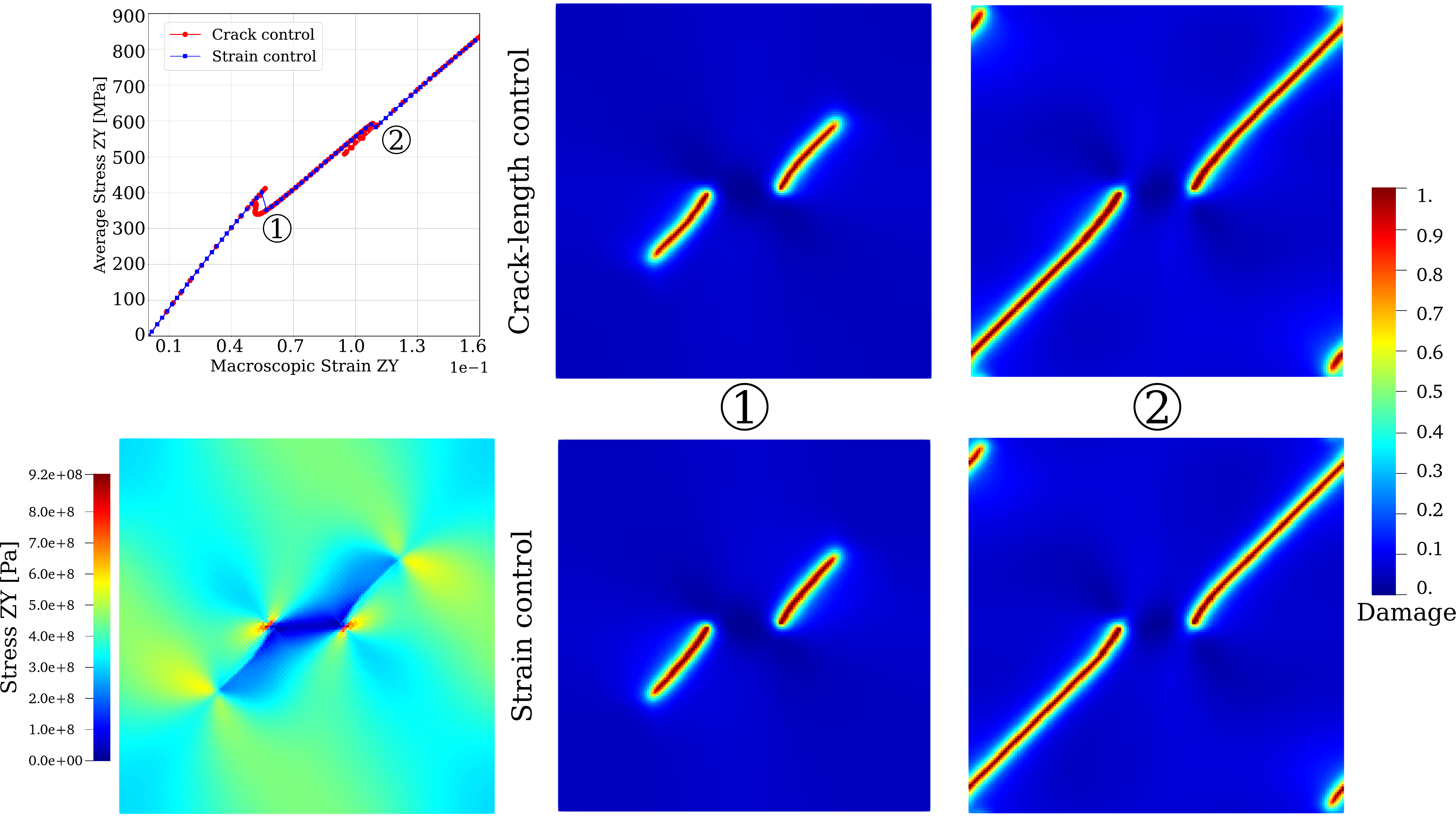}
\caption{\centering{Comparison between Crack-length and strain control models in 2D Shear simulation.}}
\label{fig6}
\end{figure}

Finally, the growth under mixed mode , as noted in section \ref{sec:strain_intro}, will be analyzed by simulating propagation under a macroscopic shear stress. 
The simulation is performed using a full periodic domain with a crack in the middle, and in this case periodicity is fulfilled in the two directions. The geometry of Fig. \ref{fig5}d is used with $A/B=1$, a initial crack length of $0.2B$ and a discretization of 243x243 voxels are used. The results are shown in Fig. \ref{fig6}, with results similar to those in Fig. \ref{fig5} for the tensile cases. It must be noted that snap-back in this configuration is much smaller being the propagation stable during most of the simulation time.


\section{Numerical examples}\label{sec:NumExamp}
In this last section of results, simulations of crack propagation in heterogeneous microstructures will be presented to show the potential of the technique developed. Two type of materials will be studied, composites and a porous material.

\subsection{Fracture of composite materials}\label{sec:fiber/laminate}
\begin{figure}[t]
\centering
\includegraphics[width=\textwidth]{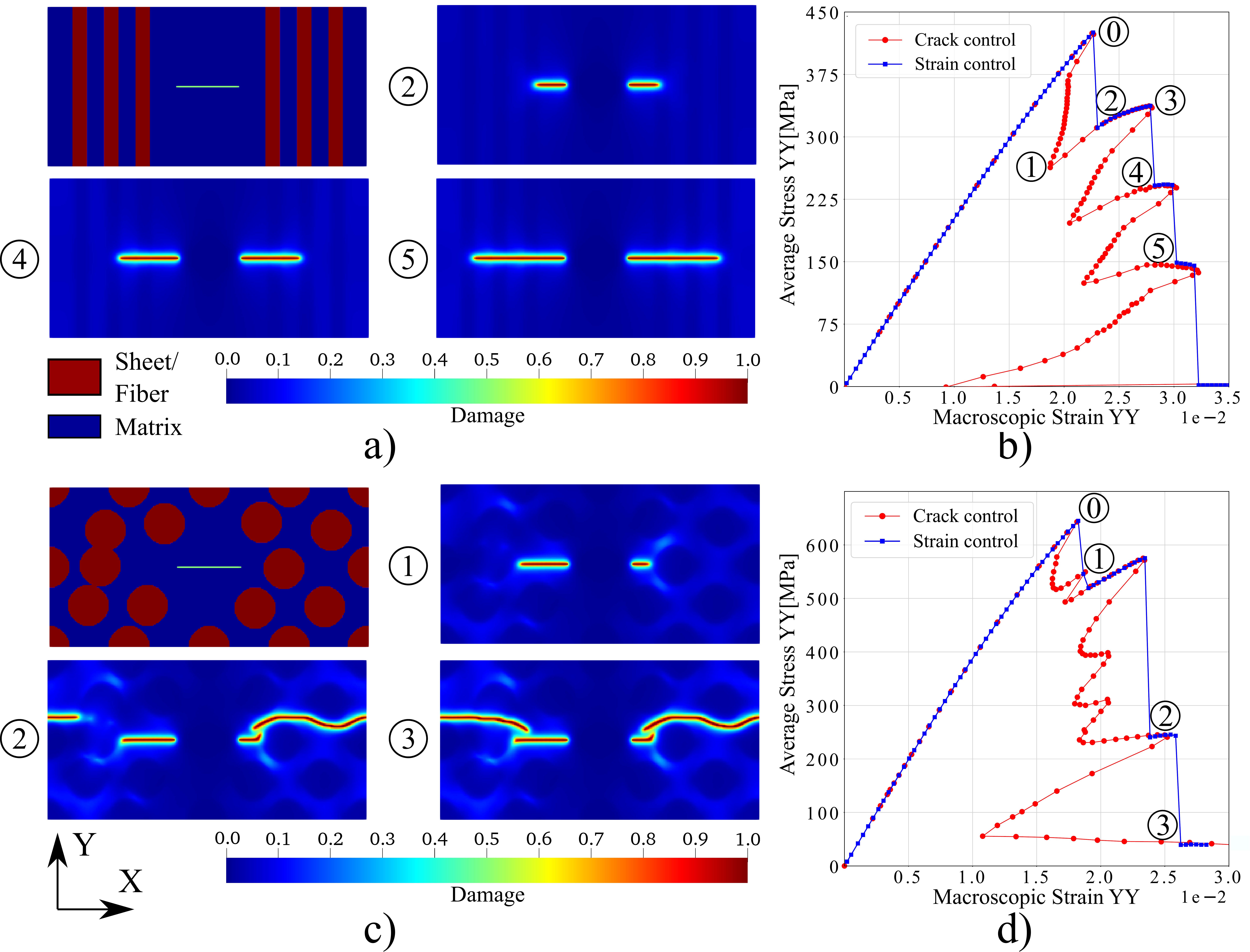}
\caption{\centering{Top: Fracture of a laminate material, (a)  crack evolution  and (b) stress-strain curve.  Bottom: Fracture of the cross section of a long fiber reinforced composite (c) crack evolution and (d) stress-strain curve.}}
\label{fig7}
\end{figure}

The following simulations correspond to the fracture of a RVE of a laminate and a fiber reinforced material under a macroscopic uniaxial tensile case (see table \ref{table:1}). Both cases have the same elastic matrix with an initial crack and no overelaxtion is used in Newton-Raphson. The first consists of a laminate with regularly distributed flexible sheets of equal thickness and perpendicular to the initial crack \bluecom{that represent an area fraction of 27.1\%}. The second case corresponds to the fracture in the transverse section of a long fiber reinforced composite with randomly distributed rigid fibers with circular cross section (see table \ref{table:2}) \bluecom{that represent an area fraction of 38.5\%}. These geometries are indeed 2D sections of 3D microstructures under plain strain. The domain shown in Fig. \ref{fig7}a is loaded by a macroscopic strain in the Y direction, parallel to the laminate and in the case of Fig. \ref{fig7}c macroscopic strain is perpendicular to the fibers. All materials are perfectly bonded and no heterogeneous crack resistance is considered in the domains, so the fracture propagation is only driven by the heterogeneity of strain microfields due to the stiffness difference between the phases. \bluecom{A unique fracture field is chosen for the whole domain, and no additional condition on the fracture field are imposed in the internal interfaces,  so cracks are allowed, if energetically favored, to penetrate either phase without any restrictions}. The RVEs are discretized in a grid of 121x243 and the result of both cases are shown in Fig. \ref{fig7}.

The curve of Fig. \ref{fig7}b shows the stress-strain behavior obtained \bluecom{for the layered composite} under the two type of controls presented. The curve obtained using crack control present a saw-tooth shape in which each snap back corresponds with the fracture of a different layer of the geometry. The curve obtained under strain control present several jumps, which corresponds to the regions of unstable growth. Different stages have been marked with numbers that correspond with the damage fields in Fig. \ref{fig7}a. By the parallel disposition of the layers, the matrix tend to accumulate more potential energy than the flexible sheets due to the difference of stiffness. When the crack start at $\tikz[baseline=(char.base)]{\node[shape=circle,draw,inner sep=2pt] (char) {0};}$ in Fig. \ref{fig7}b, the matrix is the first to be broken. The local strain caused by the crack allows the near sheet to accumulate potential energy, but this accumulation is lower than the loss of energy in the matrix, so a similar fracture process than the example of Fig. \ref{fig5}a occurs between points $\tikz[baseline=(char.base)]{\node[shape=circle,draw,inner sep=2pt] (char) {0};}$ and $\tikz[baseline=(char.base)]{\node[shape=circle,draw,inner sep=2pt] (char) {1};}$ of Fig. \ref{fig7}b where the macroscopic strain and stress decrease. Then, the fissure enters to the first sheet where the crack can grow stably because the next section of matrix accumulates more energy than is lost in the sheet. This allows the macroscopic strain and stress between points $\tikz[baseline=(char.base)]{\node[shape=circle,draw,inner sep=2pt] (char) {1};}$ and $\tikz[baseline=(char.base)]{\node[shape=circle,draw,inner sep=2pt] (char) {3};}$ of Fig. \ref{fig7}b to increase. The stable crack growth allows to capture the process in the strain-controlled simulation between points $\tikz[baseline=(char.base)]{\node[shape=circle,draw,inner sep=2pt] (char) {2};}$ and $\tikz[baseline=(char.base)]{\node[shape=circle,draw,inner sep=2pt] (char) {3};}$ of Fig. \ref{fig7}b, validating the results obtained during the staggered iterations. Similar processes occurs for the other two ramifications and the damage state of points $\tikz[baseline=(char.base)]{\node[shape=circle,draw,inner sep=2pt] (char) {2};}$, $\tikz[baseline=(char.base)]{\node[shape=circle,draw,inner sep=2pt] (char) {4};}$ and $\tikz[baseline=(char.base)]{\node[shape=circle,draw,inner sep=2pt] (char) {5};}$ of Fig. \ref{fig7}b are shown in Fig. \ref{fig7}a, where it can be seen that the crack is in the middle of each flexible sheet.

Similar behavior is observed in the curves of Fig. \ref{fig7}d for the fiber reinforced material where, due to the random distribution of the fibers, the ramifications are less symmetric than in the laminate. Since the fibers are several times stiffer than the matrix, they are not able to get strain enough to accumulate the necessary energy to create cracks. This force the crack to develop mostly in the areas of the matrix close to a fiber, being the crack unable to penetrate the fibers. It can be seen in Fig. \ref{fig7}c how the crack follows a path that avoids the position of the fibers, which occurs between points $\tikz[baseline=(char.base)]{\node[shape=circle,draw,inner sep=2pt] (char) {1};}$ and $\tikz[baseline=(char.base)]{\node[shape=circle,draw,inner sep=2pt] (char) {2};}$ of Fig. \ref{fig7}d. Then, as it is shown in Fig. \ref{fig7}c for the point $\tikz[baseline=(char.base)]{\node[shape=circle,draw,inner sep=2pt] (char) {2};}$, the only resisting material is a joint group of fibers. Only in this point, the crack enters the fiber to end the fracture process between points $\tikz[baseline=(char.base)]{\node[shape=circle,draw,inner sep=2pt] (char) {2};}$ and $\tikz[baseline=(char.base)]{\node[shape=circle,draw,inner sep=2pt] (char) {3};}$ of Fig. \ref{fig7}d. Similar effect between the crack-length and strain controlled simulations is observed in both simulations. It can be also remarked that crack paths and stress-strain curves coincide with the ones obtained in a staggered approach if a very restrictive tolerance is used. In this case, several hundreds of iterations are needed to resolve the unstable stages. On the contrary, if the staggered solver is used with a less restrictive tolerance to improve convergence, differences are found both in stress-strain curve and crack shapes after instability.

\bluecom{
\subsection{Effective toughness estimation on fiber reinforced composites }\label{sec:toughnetization}}
\bluecom{
In the last section, a good agreement was reported between the value of $G_{ceff_{o}}$ obtained using the J-integral and the one obtained with the method proposed in this work (section \ref{ssec:effective_Gc}) in the case of an homogeneous medium. However, for the heterogeneous media the use of a J-integral is not straightforward since this measure is not trajectory-independent and a lot of disperse damage is present in the microstructure as it can be seen in Fig. \ref{fig7}. Moreover, some cases presented two cracks propagating at the same time, which makes difficult to apply the J-integral. For this reason the method proposed here, which only require computing the dissipated energy which makes it unaffected by the before mentioned issues, is used to analyze the change in the effective toughness of composites. Results are compared with values obtained in \cite{schneider2020fft} using a path optimization approach. Note that there are fundamental differences between the  method used in \cite{schneider2020fft} and the definition of the effective toughness proposed here, so comparisons are merely qualitative. 

The study will focus on long fiber reinforced composites in the cross section (as in Fig. \ref{fig7}c). The RVEs used consist in a homogeneous matrix with a periodic distribution of monodisperse circular inclusions. As in \cite{schneider2020fft}, the phases will have the same stiffness and a difference of $G_c$ (1:10). The domain is defined with a resolution of 121x243 voxels and the particle positions are generated with the Random Sequential Adsortion method (RSA) proposed in \cite{segurado2002numerical}. The number of inclusions and cell size is kept constant for all the volume fractions, and the fiber radius will be changed to achieve 10\%, 20\%, 35\% and 48\% of inclusion fraction.

An initial crack of 5\% of the RVE width is considered to be consistent with linear elastic fracture mechanics and PFF model, which do not consider nucleation. Therefore, the $G_{Ceff}$ values obtained with this method might not correspond with the minimal energy crack-path as estimated in \cite{schneider2020fft}, but might be influenced by the presence of the initial crack and could result in higher values of toughness for small RVEs. It is expected that a progressive increase of the RVE size will reach to an scenario where the results become insensitive to the position of the initial crack, but this study is out of the scope of this paper. To limit the dependency with the initial crack length, the $G_{Ceff}$ obtained for a composite are normalized with the toughness obtained from the fracture of the matrix without inclusions with the same $\boldsymbol{f}$ and same initial crack of the composites.

 Crack direction at the microscale  cannot be set explicitly as in \cite{schneider2020fft} and is here a consequence of the orientation of the initial microcrack and the macroscopic strain path defined through $\boldsymbol{f}$. In this study only horizontal cracks will be promoted by using configurations with $\boldsymbol{f}_{ij=11}=1$ and $\boldsymbol{f}_{ij\neq11}=0$ that, for the initial crack orientation, correspond to a mode I macroscopic crack.
 

The \emph{isotropic} version of the PFF model \cite{MieheIJNME2010} will be considered, in which the elastic free energy is $\psi=\psi_o^+ + \psi_o^-$ in Eq.\eqref{eq:SystemDisc}. This choice is made to avoid residual load transfer after cracking in localized areas when degradation is made dependent on positive/negative split of the strain tensor.
 

The results on this simulations are represented in Fig. \ref{fig7.5}, together with the results of fiber reinforced composites reported in \cite{schneider2020fft}.
\begin{figure}[htbp]
\centering
\includegraphics[width=\textwidth]{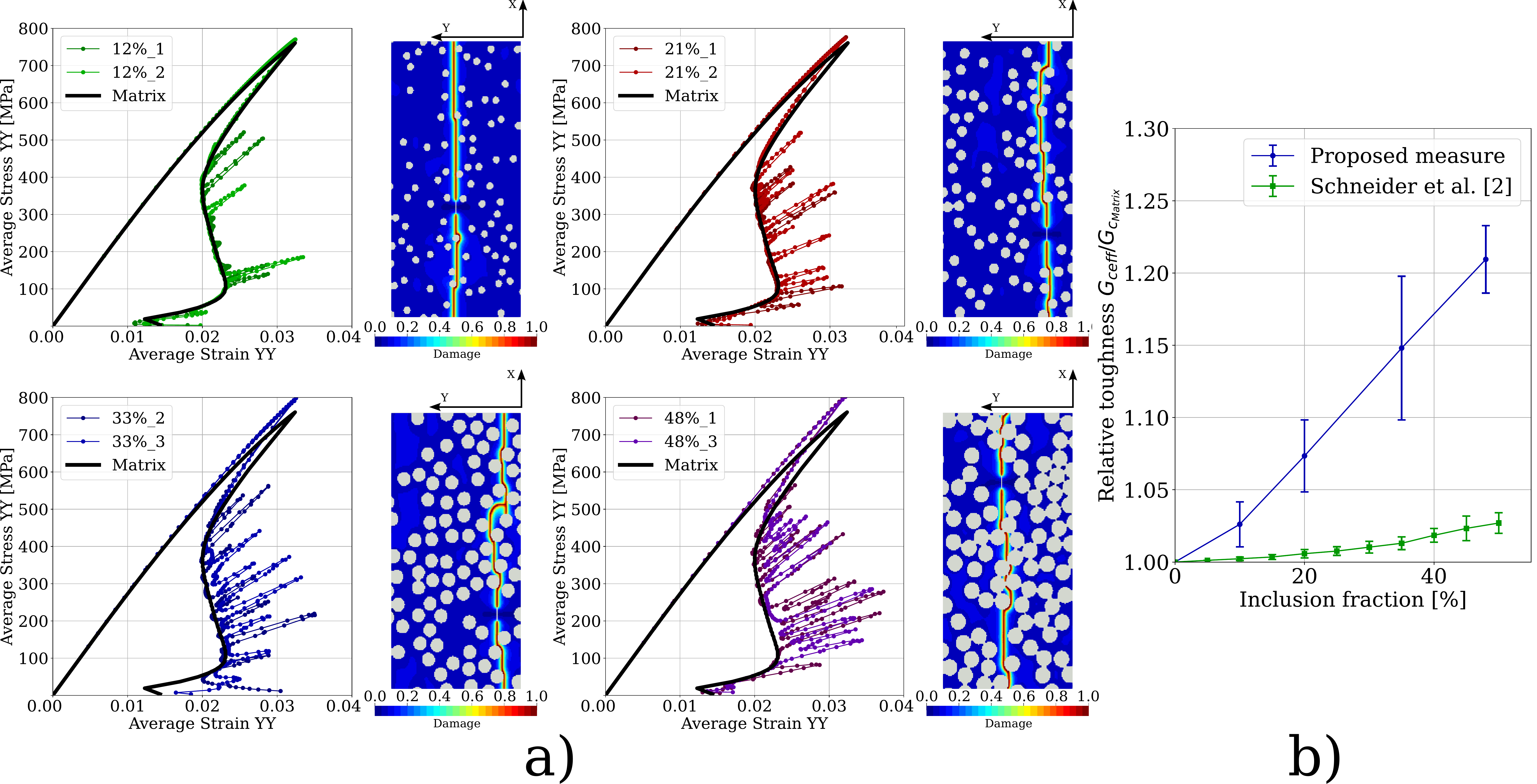}
\caption{\centering{a) Stress-strain curves and final cracks in simulations of fiber reinforced composites. b) Effective toughness as function of fiber volume fraction with current method (section \ref{ssec:effective_Gc}) and results from \cite{schneider2020fft}}}
\label{fig7.5}
\end{figure}
Although qualitatively both measures of $G_{Ceff}$ provide the same response, a non-linear toughening with increasing the volume fraction, big differences between both measures are observed, which can be explained by the inherent difference between models. First, the existence of an initial crack in our approach biases the crack to a path that might not correspond to the minimum path that would have been found in its absence. In addition, the model allows for the formation of multiple cracks which dissipate energy in their initiation, even though they did not finally propagate, which alters the measurement. 

The mechanical behavior and final crack of two selected cases in every group is shown in Fig. \ref{fig7.5}a. In the stress-strain plots, the evolution of the case with a non-reinforced matrix is included. It can be seen that all cases present loading peaks in the stress-strain response which unloading stage lies exactly on the matrix response. These peaks, similar to the ones the reported in \cite{hossain2014effective}, increase the are under the curve and are evidence of the extra energy required to fracture the microstructure. Since all cases are normalized by the same macroscopic crack length $L_g$ of Eq.\eqref{eq:JaviGc}, the greater the presence of fibers, the more peaks in the response and therefore the larger the value of $G_{Ceff}$ will be. Note that crack path tortuosity correspondingly increases with volume fraction, since in a Gamma-convergency analysis the dissipated energy should equal the integral of the fracture energy following the crack path.}

\subsection{3D cases: Circumferential fissure and porous material} \label{sec:3Dfisure/pore}
\begin{figure}[htbp]
\centering
\includegraphics[width=\textwidth]{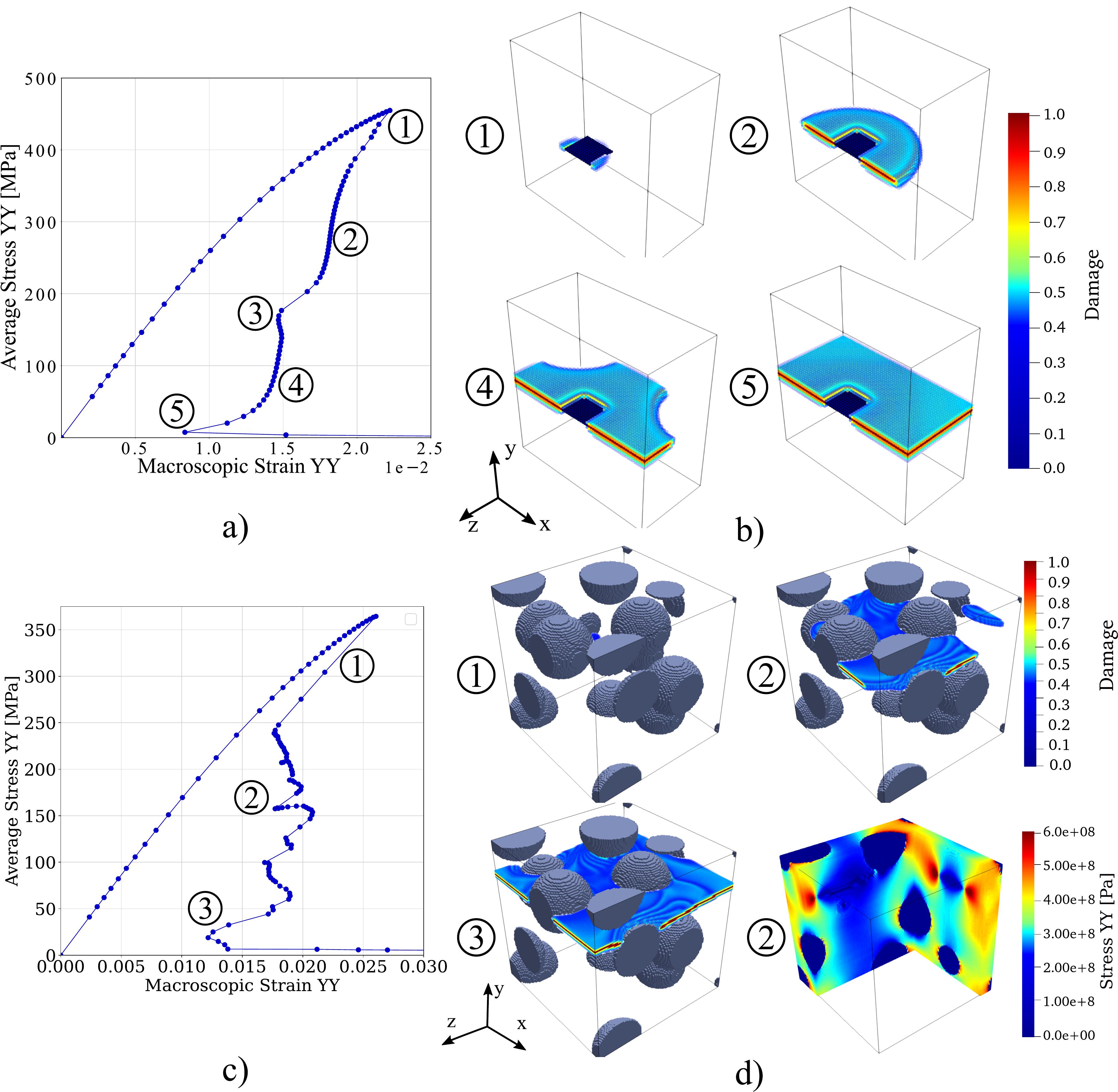}
\caption{\centering{Simulations of two 3D cases with Crack-length control. Top: homogeneous material with a single rectangular initial crack, (a) Stress-strain curve, (b) crack evolution. Bottom: Porous material with randomly distributed spherical pores, (c) Stress-strain curve, (d) crack evolution.}}
\label{fig8}
\end{figure}

In order to show the validity and efficiency of the method proposed, full three dimensional microstructures are simulated. \bluecom{Two cases are studied. The first one corresponds to an homogeneous elastic matrix with a rectangular crack in the center. The second case is a porous material, represented as a matrix with an array of 8 randomly distributed spherical voids (material properties in table \ref{table:2}) with a void fraction of 20\%. The RVE is discretized in 105x105x105 voxels. The boundary conditions described at the beginning of section \ref{sec:results} are also used here.} Simulations are performed using a crack surface control and the non-linear resolution of the system (Eq. \ref{eq:System_CrackLength} ) includes here a relaxation parameter of $\alpha=0.9$, since it was found this value improved the convergence. Simulations results are shown in Fig. \ref{fig8}.

Two stages can be identified in Fig. \ref{fig8}a for the first case, where the first stage covers points $\tikz[baseline=(char.base)]{\node[shape=circle,draw,inner sep=2pt] (char) {1};}$ to $\tikz[baseline=(char.base)]{\node[shape=circle,draw,inner sep=2pt] (char) {3};}$ and the second stage covers points $\tikz[baseline=(char.base)]{\node[shape=circle,draw,inner sep=2pt] (char) {3};}$ to $\tikz[baseline=(char.base)]{\node[shape=circle,draw,inner sep=2pt] (char) {5};}$. Their damage states are shown in Fig. \ref{fig8}b, where thanks to the symmetry only the half of the domain is presented to show the middle YX plane for a better understanding. From $\tikz[baseline=(char.base)]{\node[shape=circle,draw,inner sep=2pt] (char) {1};}$ to $\tikz[baseline=(char.base)]{\node[shape=circle,draw,inner sep=2pt] (char) {3};}$, the snap-back behavior is similar to that of the previous sections, but the crack propagates in the two directions of the plane, transforming from a square crack to an ellipsoidal one, as  observed in Fig. \ref{fig8}b for the point $\tikz[baseline=(char.base)]{\node[shape=circle,draw,inner sep=2pt] (char) {2};}$. After point $\tikz[baseline=(char.base)]{\node[shape=circle,draw,inner sep=2pt] (char) {2};}$, crack growth becomes anisotropic 
and accelerates in the areas near the periodic boundaries due to the interaction with the periodic cracks. Point $\tikz[baseline=(char.base)]{\node[shape=circle,draw,inner sep=2pt] (char) {3};}$ in Fig. \ref{fig8}a represents the coalescence of the crack, which alters the stress-strain path due to the abrupt change in stiffness that occurs in the matrix before both crack fronts encounters. In the second stage the rest of the remaining material is fractured in a similar process until the point $\tikz[baseline=(char.base)]{\node[shape=circle,draw,inner sep=2pt] (char) {5};}$ where the fracture is complete.

In the case of the porous material, the stress-strain response is shown in Fig. \ref{fig8}c with damages states \bluecom{and stress states} shown in Fig. \ref{fig8}d for the points marked in the graphic. No initial sharp fissure was introduced, and the crack propagates from the initial voids. \bluecom{The initial crack propagation} (point $\tikz[baseline=(char.base)]{\node[shape=circle,draw,inner sep=2pt] (char) {1};}$) occurs due to the finite value of $\ell$, and no crack would appear for $\ell\rightarrow\infty$ since linear elastic fracture mechanics just considers propagation from initial sharp discontinuities. \bluecom{The position at which the crack nucleates (point $\tikz[baseline=(char.base)]{\node[shape=circle,draw,inner sep=2pt] (char) {1};}$ of the stress-strain curve) is shown of Fig. \ref{fig8}d, where the damage field is represented.} Then, the crack grows following a similar process to those studied in previous sections and is shown in Fig. \ref{fig8}d for the points $\tikz[baseline=(char.base)]{\node[shape=circle,draw,inner sep=2pt] (char) {2};}$ and $\tikz[baseline=(char.base)]{\node[shape=circle,draw,inner sep=2pt] (char) {3};}$. In this case, the stress field in Y direction is presented for the point $\tikz[baseline=(char.base)]{\node[shape=circle,draw,inner sep=2pt] (char) {2};}$ in Fig. \ref{fig8}d, in a plane that allows to show 3 different crack fronts. It can be seen that the stress concentration fields around the crack are similar with those presented in Fig. \ref{fig3_b} for 2D cases.

\section{Conclusions}
A crack-growth control technique is proposed for the simulation of crack propagation at the microscale in phase field fracture and is adapted for both Finite Element and Fast Fourier Transform based solvers. In the case of FFT, this type control technique has not been proposed before due to the inherent difficulties of coupling this type of control in the iterative nature of FFT algorithms. The method allows using a monolithic scheme, avoiding the non-convexity of energy functional that causes numerical problems and taking advantage of the efficiency of this numerical framework. \bluecom{ It is shown that the implementation proposed for FFT and FEM are equivalent and both methods produce same crack paths and very close stress-strain responses, being the response of both methods superposed for sufficiently small discretizations.}

The method allows to obtain the overall behavior including all the snap-back during crack propagation through the microstructure, showing the stable and unstable growth phases. J-integral calculations show that the control proposed is equivalent to maintaining the energy release rate constant for every state of crack growth. \bluecom{It is also shown that for an homogeneous material the critical energy release rate numerically obtained (effective numerical toughness) is slightly different to the material one, as deduced and quantified in \cite{bourdin2008variational} for FEM. This difference is the result of the particular choice of characteristic length and discretization, and the value of this effective numerical fracture energy for FFT simulations is provided.}

\bluecom{The crack control method allows to obtain an estimation of the homogenized toughness of heterogeneous materials. This can be done by obtaining the maximum of the J-integral during the crack propagation, as proposed in  \cite{hossain2014effective}, or by computing the total dissipated energy after crack complete coalescence. This last procedure has some similarities with the minimization path method proposed in \cite{schneider2020fft}, but in the present case an initial crack is required ---which influences the crack path--- and results couple mechanical and fracture fields at the microscale. The second method have been used to provide estimations of fracture toughness of long fiber reinforced composite in the cross section. The results show an increase of fracture toughness with the fiber volume fraction, in agreement with previous studies. However, the quantitative results of the effective toughness differ due to inherent difference in the definition of the macroscopic toughness estimators.}

\bluecom{As final remark, the numerical examples show the potential of the technique developed to study crack propagation at the microscopic level in complex three dimensional heterogeneous microstructures.}

\section*{Acknowledgments}
P. Aranda is grateful for the support of the doctoral program ANID-BECAS CHILE 2020-72210273 from the Government of Chile and to the Vicerrectoría de Investigación, Desarrollo e Innovación of the Universidad de Santiago de Chile. We gratefully acknowledge the funding of the Universidad Politécnica de Madrid through project REM20431JSE.


\clearpage

\section{Appendix}\label{ApendSec}
\subsection{Effective critical energy release rate comparison}\label{EffGcsec}
As was exposed in section \ref{sec:stability}, values of energy release rate measured with the J-integral are not accurate respect the imposed critical energy release rate (called from now the imposed toughness), but an overestimation is always obtained. 

In the seminal work of Bourdin et al \cite{bourdin2008variational}, the correct representation of the energy functional in the presence of a PFF crack is studied and how this representation works when the Ambrosio and Tortorelli crack functional $\Gamma$-converge to a real crack when a FEM projection is used in a fracture problem. One conclusion of this work is the existence of an effective toughness $G_{Ceff_o}$ depending on the regularization parameter $\emph{l}$ and the size of the FEM element $h$ in the discretization:

\begin{equation}
\label{eq:AP2_BourdinAprox}
G_{Ceff_o} = G_{c} \left(1+\frac{h}{2\emph{l}} \right).
\end{equation}

This relation was probed right in many works \cite{hossain2014effective, tanne2018crack, costa2023formulations} for FEM-PFF discretizations. Nevertheless, this relation seem not to be found for the present FFT implementation considering h as voxels size. For this reason, a numerical study on the energy functional is made to check if the FFT proposal $\Gamma$-converges in the same way as in FEM. For this purpose, the dissipative part of the energy functional, defined by the formulation $\int_{\Omega}G_{c}\gamma_{f} d\Omega$ with the functional of Eq.\eqref{eq:CrackLenght}, is considered to build the most accurate approximation with a given FFT discretization and verify his behavior with $h$ and $\emph{l}$. 

From the procedure in the work of Alberti et al \cite{Alberti2000}, we consider the Miehe et al \cite{MieheIJNME2010} damage distribution $\phi(x)=e^{ -\frac{|x-a|}{2\emph{l}}}$ as the optimal minimizer of the problem, where $a$ is the position of the center of the crack in a trajectory parameterized with $x$ that is perpendicular to an already formed crack. This define a one-dimensional slice of the crack that span the periodic length of the domain $L$:

\begin{equation}
\label{eq:AP2_Param}
G_{c}\Gamma_{x} = \int_{0}^{L}G_{c}\gamma_{x} dx = G_{c} \int_{0}^{L} \frac{1}{2\emph{l}}\phi^{2}(x)+\frac{\emph{l}}{2} \left(\frac{d\phi}{dx}(x) \right)^{2} dx.
\end{equation}

The optimal damage distribution can be discretized in $N$ regular segments, which  define a finite number of real damage points $\phi_{n}$. The FFT projection of damage can be defined by means of a Fourier series approximation of the Miehe optimal distribution, taking account the discrete Fourier transform of the damage $\hat{\phi}_{k}$ with $N$ discrete frequencies:

\begin{equation}
\label{eq:AP2_FourierDiscretePhi}
\phi(x) \approx \sum_{k=-N/2}^{N/2} \frac{\hat{\phi}_{k}}{N} e^{ \frac{i2\pi k}{L}x},
\end{equation}

\noindent where $k$ represent the discrete Fourier frequency set. The definition of $\Gamma_{x}$ includes the derivative of damage, which definition in the discrete Fourier space is:

\begin{equation}
\label{eq:AP2_FourierDiscreteDPhi}
\frac{d\phi}{dx}(x) \approx \sum_{k=-N/2}^{N/2} \frac{\hat{\phi}_{k}}{N} \frac{i2\pi k}{L} e^{ \frac{i2\pi k}{L}x}.
\end{equation}

Replacing this definitions in Eq.\eqref{eq:AP2_Param} we obtain the following approximation by Fourier series of the crack integral:

\begin{equation}
\label{eq:AP2_Int1}
G_{c}\Gamma_{x} \approx \frac{G_{c}}{2\emph{l}}   \int_{0}^{L}    
\sum_{k=-\frac{N}{2}}^{\frac{N}{2}} \sum_{j=-\frac{N}{2}}^{\frac{N}{2}} \frac{\hat{\phi}_{k}\hat{\phi}_{j}}{N^{2}}  \left[1+\left(\frac{i2\pi kj\emph{l}}{L} \right)^{2}\right] e^{ \frac{i2\pi}{L}(k+j)x}dx,
\end{equation}

\noindent where $k$ and $j$ represent two independent frequency sets. The exponential part of the functional is the only depending on x, which implies that is the only integrand in equation \ref{eq:AP2_Int1}:

\begin{equation}
\label{eq:AP2_Int2}
\frac{G_{c}}{2\emph{l}}\sum_{k=-\frac{N}{2}}^{\frac{N}{2}} \sum_{j=-\frac{N}{2}}^{\frac{N}{2}} \frac{\hat{\phi}_{k}\hat{\phi}_{j}}{N^{2}} \left[1+\left(\frac{i2\pi kj\emph{l}}{L} \right)^{2}\right] \int_{0}^{L} e^{ \frac{i2\pi}{L}(k+j)x}dx,
\end{equation}

In order to express this integral correctly, all the combination that fulfills $k+j=0$ are separated from the sum an expressed as $1$. Since the Fourier representation of $\phi$ is a linear combination of trigonometric functions with a certain number of complete cycles and the integral is made for a complete period $x=0 \rightarrow L$, it is established that for $k+j\neq0$ the value of the integral is zero. The expression of Eq.\eqref{eq:AP2_Int2} results in:

\begin{equation}
\label{eq:AP2_Int3}
G_{c}\Gamma_{x} \approx \frac{G_{c}}{2\emph{l}}\sum_{k=-\frac{N}{2}}^{\frac{N}{2}} \frac{\hat{\phi}_{k}\hat{\phi}_{-k}}{N^{2}}  \left[1+\left(\frac{i2\pi kj\emph{l}}{L} \right)^{2}\right].
\end{equation}

This expression allows to calculate the integral 
of Eq.\eqref{eq:AP2_Param} for any Fourier projection of $\phi$, where the use of a certain $N$ define the voxel size $h_{v}$. 

As stated above, in both FEM and FFT models the imposed toughness is amplified by a factor $A$, where the first is defined in Eq.\eqref{eq:AP2_BourdinAprox}. For given $h_{v}$ and $\emph{l}$, the Fourier approximation of $A$ as well as $\phi$ and his derivative are shown in Fig. \ref{figA1}:

\begin{figure}
\centering
\includegraphics[width=130mm]{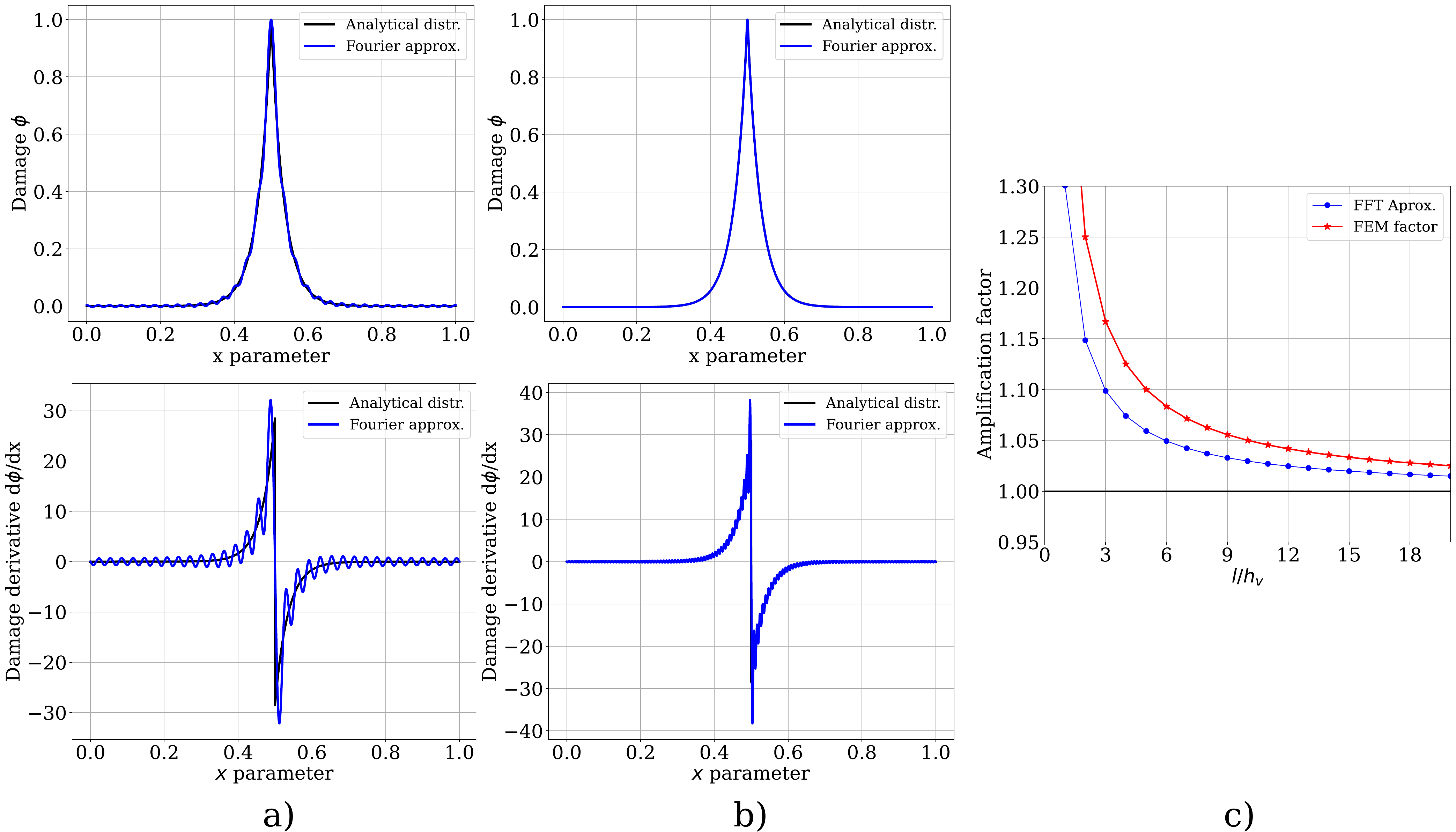}
\caption{\centering{(a and b) FFT approximation of damage distributions for different discretizations and (c) FFT and FEM deviations from imposed toughness.}}
\label{figA1}
\end{figure}

where, the cases in Fig. \ref{figA1}a and Fig. \ref{figA1}b are made for a discretization of $N=65$ and $N=257$. It can be seen how the Fourier approximation converges to the analytic distribution of damage that is already known that $\Gamma$-converges. The amplification factor $A$ is defined as the ratio $G_{Ceff_o}/G_{C}$ and approximates to $1$ when the ratio $\emph{l}/h_{v}$ grows. Its also can be seen that this occurs faster than with the FEM approximation, which can be due to the Fourier approximation of the derivative of damage, that have a great approximation of the derivative in the proximity of the center of the crack despite the oscillatory behavior observed.

\begin{figure}[ht]
\centering
\includegraphics[width=90mm]{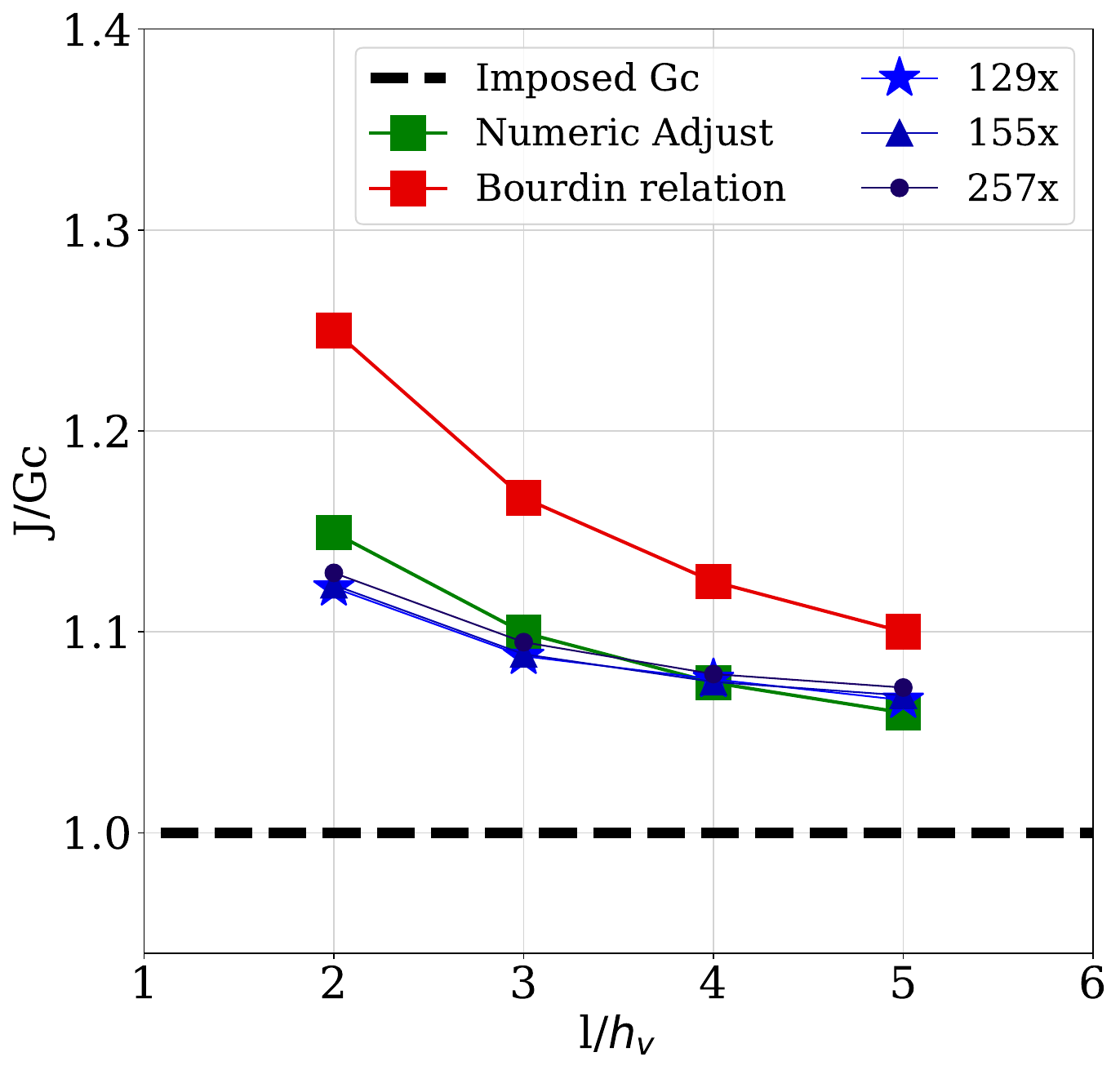}
\caption{\centering{Comparison between the aproximation of the FFT amplification factor with J-integral measures of $G_{Ceff_o}$ in real simulations of $129x129$, $155x155$ and $257x257$ voxels.}}
\label{figA2}
\end{figure}

In this work, this numerical value of the amplification factor is used to estimate $G_{Ceff_o}$. His behavior is compared with real numerical simulations, where the energy release rate is obtained with the J-integral, as can be seen in Fig. \ref{figA2}.

The comparison shows certain deviation that could be due to the history variable which was not considered in the above study and that could lead to damage distributions that are not the analytical approximation of Eq.\eqref{eq:AP2_FourierDiscretePhi}. Despite this, a good agreement between the approximation and simulation values is obtained. The numerical adjust observed in  Fig. \ref{figA2} in green, correspond to a least squares calculation over a function $f_{ad}=1+\frac{a}{\emph{l}/h_{v}}$ to resembles the Bourdin formulation. A value of $a=0.298$ is obtained which is less than the value $0.5$ postulated for FEM.

\end{document}